\documentclass[sigconf,pdfusetitle,pdfdisplaydoctitle]{acmart}
\AtBeginDocument{%
  }

\setcopyright{acmlicensed}
\copyrightyear{2026}
\acmYear{2026}
\setcopyright{cc}
\setcctype{by}
\acmConference[UIST '26]{The 39th Annual ACM Symposium on User Interface Software and Technology}{November 02--05, 2026}{Detroit, MI, USA}
\acmBooktitle{The 39th Annual ACM Symposium on User Interface Software and Technology (UIST '26), November 02--05, 2026, Detroit, MI, USA}
\acmDOI{10.1145/3830398.3830718}
\acmISBN{979-8-4007-2856-3/2026/11}

\usepackage{todonotes}
\usepackage[title]{appendix}
\usepackage{booktabs}
\usepackage{makecell}
\usepackage{enumitem}

\newcommand{\change}[1]{{#1}}
\begin{document}

\title{SemanticSlider3D: Training-Free Continuous Semantic Editing for 3D Objects}


\settopmatter{authorsperrow=2}

\author{Ru Wang}
\authornote{Work done during internship at Fujitsu Research of America. \\ Project page: \url{https://ruw001.github.io/semanticslider3d-page/}}
\email{ru.wang@wisc.edu}
\orcid{0000-0003-3907-2370}
\affiliation{%
  \department{Department of Computer Sciences}
  \institution{University of Wisconsin--Madison}
  \city{Madison}
  \state{Wisconsin}
  \country{USA}
}

\author{Rahul Jain}
\email{jain348@purdue.edu}
\orcid{0009-0001-3723-5482}
\affiliation{%
  \department{Department of Electrical and Computer Engineering}
  \institution{Purdue University}
  \city{West Lafayette}
  \state{Indiana}
  \country{USA}
}

\author{Koichiro Niinuma}
\email{kniinuma@fujitsu.com}
\orcid{0000-0001-8367-3988}
\affiliation{%
  \institution{Fujitsu Research of America}
  \city{Pittsburgh}
  \state{Pennsylvania}
  \country{USA}
}

\author{Aakar Gupta}
\email{agupta@fujitsu.com}
\orcid{0000-0001-6435-3583}
\affiliation{%
  \institution{Fujitsu Research of America}
  \city{Redmond}
  \state{Washington}
  \country{USA}
}

\renewcommand{\shortauthors}{Wang et al.}

\begin{abstract}
Fine-grained control over continuous semantic attributes of 3D objects is essential for 3D content creation, but is not well supported by conventional 3D modeling workflows or prompt-based interaction with existing generative AI tools. While slider-based methods have proven effective for fine-grained semantic control in 2D image generation, no equivalent approach exists for 3D. Extending these 2D methods to 3D is non-trivial due to challenges unique to 3D, including geometric integrity and cross-view coherence. We present SemanticSlider3D, a technique for continuous semantic attribute editing of 3D objects that requires no per-attribute training. Given a user-specified attribute, our pipeline constructs a semantic editing direction in the latent space of a state-of-the-art 3D generation model, presenting a diverse and coherent spectrum of 3D variations. A technical validation on a dataset of 50 3D object-attribute pairs shows \change{our method was preferred by all five human assessors across variation range, consistency, 3D object quality, and attribute disentanglement, over a baseline combining a 2D slider with an image-to-3D model.} An exploratory study with six participants demonstrates that SemanticSlider3D supported decision-making in 3D prototyping and was perceived as a valuable addition to existing workflows.
\end{abstract}




\begin{teaserfigure}
  \includegraphics[width=\textwidth]{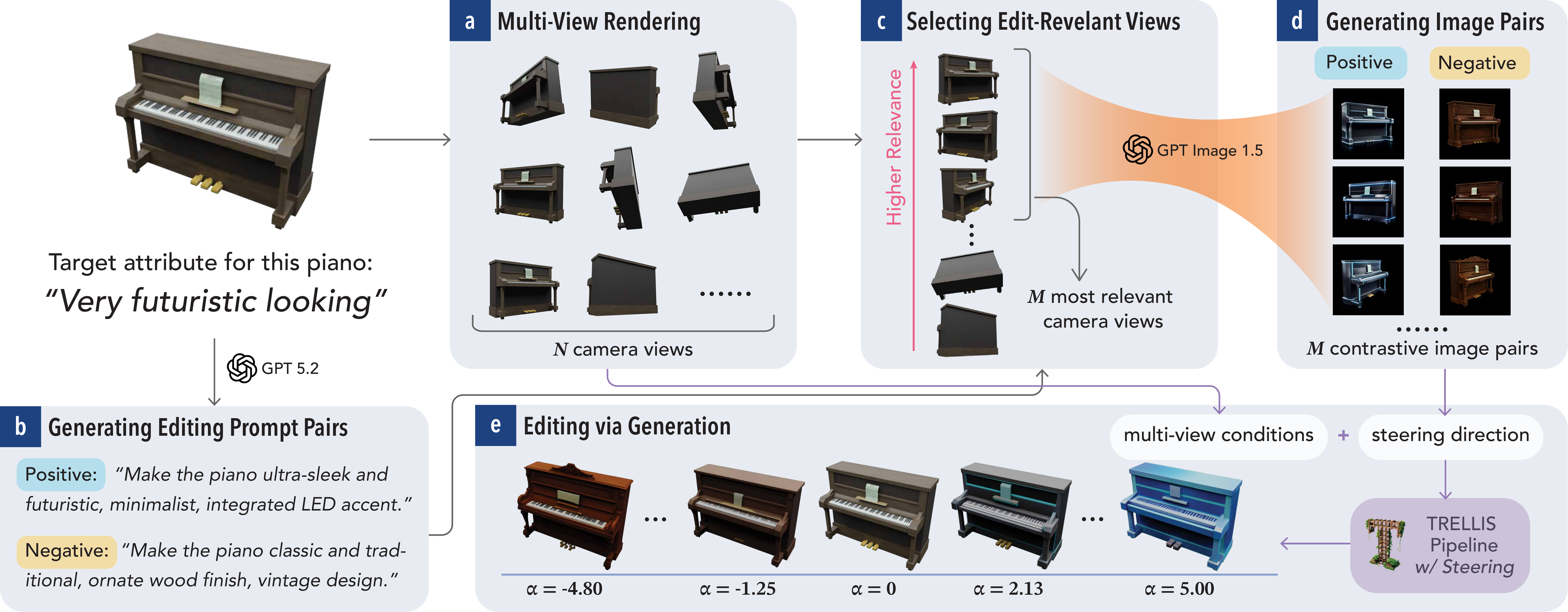}
  \caption{\change{The pipeline of generating image conditions and editing through generation process. We use a piano as an example, with "very futuristic looking" being the target attribute specified by the user. (a) Rendering $N$ camera views from the original 3D object; (b) generating contrastive editing prompts for both positive direction and negative direction based on user-specified attribute; (c) ranking and selecting camera views relevant to the editing prompts; (d) generating contrastive image pairs using edit-relevant views and editing prompt pairs for image editing; (e) Editing 3D objects along the semantic axis using TRELLIS with multi-view conditions and steering direction defined by contrastive image pairs.}}
  \label{fig:pipeline}
\end{teaserfigure}


\maketitle

\section{Introduction}
3D design is an essential activity in product design \cite{park2026generative}, architectural design \cite{bayar2018rapid}, game development \cite{hristov2021workflow}, and many other creative industries. The early stage of 3D design involves \textit{generating}, \textit{refining}, and \textit{iterating} on ideas about the appearance of 3D models \cite{park2026generative}, which can be time-consuming and requires significant manual effort with conventional 3D modeling workflow \cite{park2026generative, camburn2017design}. 

Recent advances in generative AI allow users to generate and edit high-quality 3D assets through text and image prompts \cite{li2025step1x, xiang2025structured, jun2023shap, chen2026sam, xu2024instantmesh}, and such tools have been increasingly integrated into 3D prototyping workflows \cite{edwards2024sketch2prototype, faruqi2023style2fab, vachha2025dreamcrafter, lee2025imaginatear, shen2025gaussianshopvr}. Through prompt-based interactions, users can rapidly create diverse alternatives and edit 3D objects to refine prototypes. However, these generative models largely operate in an end-to-end fashion, making it difficult to intervene between the prompt and the output. This is particularly problematic for editing continuous semantic attributes such as style or material finish \cite{gandikota2024concept, ezra2025freesliders, chung2023promptpaint}, where users need fine-grained, incremental control over the intensity of the edit rather than a binary change \cite{chung2023promptpaint, duan2025designfromx}. In practice, this leads to repetitive prompting until a generation aligns with the user's expectation, and regeneration often changes unintended attributes \cite{edwards2024sketch2prototype}.


There has been extensive prior work on continuous control of semantic attributes for 2D images using slider-based interfaces \cite{ezra2025freesliders, gandikota2024concept, dang2022ganslider, evirgen2023ganravel, guerrero2024texsliders, ekin2026unreasonable, chung2023promptpaint}. Although slider-based editing has proven effective for precise image generation and editing, it remains unclear how this approach applies to 3D, given challenges specific to 3D such as structural integrity \cite{edwards2024sketch2prototype} and multi-view consistency \cite{shen2025gaussianshopvr, poole2022dreamfusion, tang2024lgm}.


To close this gap, we propose \textit{SemanticSlider3D}, a slider-based semantic attribute editing technique for 3D objects that requires no per-attribute training. With SemanticSlider3D, the user specifies a continuous attribute for any 3D object using a text prompt. Our pipeline selects the most representative camera views relative to the target attribute, generates contrastive image pairs aligned with the negative and positive extremes of the attribute, and leverages TRELLIS \cite{xiang2025structured}, a state-of-the-art image-to-3D model, to create an editing direction in the latent space. At inference time, the user sees multiple variations on a slider representing different intensities of the edit in both directions. Users can add intermediate points for finer granularity and perform sequential editing by anchoring new sliders on existing variants.


To validate our technique, we conducted a technical evaluation with five human assessors using a custom 3D dataset of diverse everyday objects and well-defined continuous attributes. 
We found that SemanticSlider3D demonstrated high variation range, reasonable variation consistency, good 3D object quality, and strong preservation of unintended attributes, and was overwhelmingly preferred by human assessors over a baseline combining a state-of-the-art image slider \cite{gandikota2024concept} with a state-of-the-art image-to-3D model \cite{chen2026sam}.
We further conducted an exploratory user study with six participants of varying 3D prototyping experience to understand whether and how slider-based 3D editing supports 3D prototyping. We found that SemanticSlider3D facilitated decision-making in exploring and refining design ideas, helped communicate the generative AI model's capabilities, and was perceived as a useful addition to participants' 3D prototyping workflows.  

Our contributions are three-fold: (1) a novel slider-based 3D semantic editing technique; (2) a technical validation with human assessors on a representative custom dataset, demonstrating the effectiveness of our technique; and (3) an exploratory user study showing the potential of SemanticSlider3D for 3D prototyping.


\section{Related Work}


\subsection{3D Generation and Editing}

\subsubsection{3D Generation}
Recent advances in computer vision have enabled the creation of 3D content from text, images, and other modalities \change{\cite{xiang2025structured, mildenhall2021nerf, poole2022dreamfusion, kerbl20233d, wang2023prolificdreamer, li2025voxhammer, jun2023shap}}. One direction is to directly reconstruct or generate 3D content. \change{Early approaches relied on Neural Radiance Fields (NeRF) and optimization-based pipelines to reconstruct 3D objects/scenes from sparse inputs \cite{mildenhall2021nerf, pumarola2021d, hong2023lrm}. Recent work has explored diffusion-based models for 3D generation \cite{poole2022dreamfusion, lin2023magic3d}, autoregressive approaches inspired by GPT-style modeling for 3D synthesis \cite{chen2024meshanything}, and efficient representations such as 3D Gaussian Splatting (GS) for real-time rendering and editing \cite{kerbl20233d}. More recently, rectified flow models have been explored as an alternative to diffusion-based methods and have shown strong performance for large-scale 3D generation \cite{xiang2025structured, chen2026sam}. Another direction is to leverage 2D image generation models for 3D generation, due to their strong generalization abilities. These approaches typically optimize 3D representations by distilling knowledge from pretrained image diffusion models \cite{wang2023prolificdreamer, tang2023dreamgaussian, liang2024luciddreamer, lin2023magic3d}. Alternatively, some methods first generate multi-view images using diffusion models and then reconstruct the corresponding 3D assets from these views\cite{li2023instant3d, xu2024instantmesh, shi2023mvdream}. However, since these methods rely on 2D generation, they often result in weaker geometry compared to models trained on 3D data, mainly due to inconsistencies across different views.}

\subsubsection{3D Editing}

Most existing 3D editing approaches are prompt-based and rely on 2D representations as intermediaries, either optimizing 3D representations via distillation from pretrained diffusion models \cite{chen2024shap, palandra2024gsedit, sella2023vox} or editing multi-view images and reconstructing the 3D assets iteratively \change{\cite{barda2025instant3dit, chen2024generic, erkocc2025preditor3d, haque2023instruct}}. However, these methods suffer from multi-view inconsistencies. A few works explore directly editing 3D latent representations \change{\cite{achlioptas2023shapetalk, wang2025garmentcrafter, chen2024gaussianeditor, parelli20263d}}, However, these methods offer limited control, because edits are specified through high-level prompts without explicit control over the magnitude or structural consistency of the change. Rectified flow-based models \cite{li2025triposg, xiang2025structured} provide a more promising foundation by directly modeling 3D data distributions with improved control for structured editing. However, like prior approaches, TRELLIS relies on discrete text prompts and does not support continuous control over semantic attributes. 
\change{Avatar Concept Slider \cite{foo2024avatar} brings continuous slider-based control to 3D, but is specific to human avatars and requires per-concept training on a 3D GS representation.
No existing method supports training-free continuous semantic editing for general 3D objects. SemanticSlider3D builds on TRELLIS by introducing a pipeline for precise and continuous 3D editing.}

\subsection{3D generative authoring systems in HCI}

3D generation and editing have been explored in HCI to support design and creative workflows, 3D authoring, and 3D content creation \cite{choi2025genpara, shen2025gaussianshopvr, duan2025designfromx, 10.1145/3742413.3789142, liu20233dall,faruqi2023style2fab}. Systems such as DreamCrafter \cite{vachha2025dreamcrafter} and GaussianShopVR \cite{shen2025gaussianshopvr} enable users to generate and modify 3D content using 3D generation models with text prompts and immersive interfaces. For example, GaussianShopVR allows point-level editing of 3D Gaussian Splatting scenes, combining manual selection tools with AI-driven editing to provide fine-grained control. Other works integrate 2D generative models into 3D design pipelines, such as combining 3D sketching with image generation or incorporating text-to-image models into 3D workflows (e.g., 3DALL-E \cite{liu20233dall}). DesignFromX \cite{duan2025designfromx} further explores user-driven design by allowing the composition of features from images of reference products. Other work has explored direct manipulation and sketch-based interfaces for 3D modeling \cite{yu2022piecewise} and AR/VR-based interaction systems that enable users to interact with 3D content in immersive environments \cite{shen2025gaussianshopvr, jiang2024vr}. 
However, most of these systems rely on image or prompt-based interactions, where users specify edits at a high level but have limited ability to continuously control or refine specific attributes. This makes it difficult to perform fine-grained adjustments or maintain consistent control over spatial and semantic properties of the 3D scene. In this work, SemanticSlider3D address this limitation by enabling continuous control in 3D editing.

\subsection{Semantic Sliders for 2D Image Editing}

Slider-based editing of semantic attributes for 2D images has been explored in multiple recent works due to its ability to provide user-friendly, low-dimensional, continuous control over semantic attributes \cite{jain2025adaptivesliders, dang2022ganslider, evirgen2023ganravel, evirgen2022ganzilla}. These methods typically learn or identify semantic directions in the latent space, either using lightweight training (e.g., LoRA adapters) or training-free approaches that leverage pretrained models. Moving along these directions enables continuous control over attributes. Surprisingly, no such methods exist for 3D objects. This is partly because semantic attributes are easier to define and control in 2D images: \change{Many pretrained generative models provide well-aligned latent directions that map to visual semantics.
In contrast, 3D representations generally lack such disentangled semantic axes.} Edits also need to preserve geometric structure and remain consistent across multiple views.
This makes continuous control in 3D significantly more challenging. An obvious approach for 3D sliders would be to leverage 2D sliders for 3D editing by projecting 3D content into a 2D representation, applying semantic edits in the image domain, and mapping the results back to 3D. 
Such 3D-2D-3D pipelines are appealing as they reuse well-established editing techniques without requiring native 3D control mechanisms. We explored this pipeline as a baseline. 
\change{However, we found that although this is a good starting point, the edits do not consistently translate to the intended 3D outcomes. Because each stage of this pipeline can introduce fidelity loss, making it difficult to enforce spatial relationships and physical constraints across the full pipeline. Thus, directly extending 2D slider approaches to 3D settings is insufficient. We propose SemanticSlider3D, which performs semantic editing directly in the 3D latent space to avoid this compounding fidelity loss.}
\section{SemanticSlider3D: Design Goals}
Given a 3D object and a user-specified semantic attribute with a continuous range of variation, our goal is to generate variations at different positions along the full spectrum defined by that attribute, for example, a steak at different levels of doneness. To enable continuous semantic editing for 3D objects while ensuring a reasonable and effective slider-based interaction, we incorporate insights from prior literature in both 3D generation \cite{xiang2025structured, xu2024instantmesh, shi2023mvdream, edwards2024sketch2prototype} and 2D image sliders \cite{gandikota2024concept, jain2025adaptivesliders, ekin2026unreasonable, evirgen2022ganzilla, evirgen2023ganravel}, and derive the following design goals (DG) for SemanticSlider3D:

\textbf{\textit{DG1: Ensuring 3D Coherence.}} 
Variations generated along an semantic attribute should be visually coherent and plausible. Unlike 2D images, 3D objects involve both geometry and appearance, meaning directly lifting image variations from a 2D slider to 3D can be problematic because important structural information may be lost or inconsistent across camera views \cite{shi2023mvdream}. Even with multi-view 3D generation models, coherence issues, such as the Janus problem (3D objects with multiple faces or heads), can arise in the resulting edited objects \cite{chen2024generic}. Prior work has emphasized the importance of maintaining geometric and appearance quality in generated 3D objects \cite{edwards2024sketch2prototype, vachha2025dreamcrafter, shen2025gaussianshopvr}. We aim to ensure that each variation along the slider preserves the overall structural integrity of the original object while applying the intended edit consistently across views.

\textbf{\textit{DG2: Wide Range of Meaningful Variations.}} 
A wide range of variations helps users explore how the attribute manifests at different intensities and identify the level that best matches their intent \cite{jain2025adaptivesliders}. Drawing insights from 2D slider-based editing, the effective range of a slider varies depending on the specific object and attribute---some combinations produce visible changes within a narrow range while others require larger shifts before the edit becomes visually apparent \cite{jain2025adaptivesliders, gandikota2024concept}. 
Therefore, the slider bounds should be adapted for each object and attribute while maximizing the diversity between the two ends.

\textbf{\textit{DG3: Consistent Variation within the Slider.}} Variations along the slider should progress smoothly and predictably, so that each step produces visually proportional change in the attribute. A challenge with slider-based editing is that the latent space can be highly nonlinear. Users may see incremental changes across several positions and then see an abrupt jump at another\cite{jain2025adaptivesliders, ezra2025freesliders, ekin2026unreasonable}, leading to expectation mismatch \cite{jain2025adaptivesliders}. We aim to ensure that SemanticSlider3D produces variations that change gradually and uniformly across the slider range, so that moving the slider by a similar amount results in a similar degree of visual change.

\textbf{\textit{DG4: Preserving Unintended Attributes.}} Editing one attribute should not change other irrelevant attributes of the 3D object. Prior work in 2D slider has proposed various methods to isolate the target edit direction from unrelated attributes (i.e. disentanglement)\cite{gandikota2024concept, ezra2025freesliders, ekin2026unreasonable}. We aim to ensure that SemanticSlider3D modifies only the intended attribute while preserving the remaining characteristics of the original 3D object as much as possible.

\change{DG2–DG4 extend challenges from 2D sliders, but manifest differently in 3D (\S\ref{method}). DG1 is unique to 3D and is a critical challenge to solve for 3D semantic sliders.}
\section{SemanticSlider3D: Design \& Implementation}
\label{method}
Building on these design goals, we present the pipeline of SemanticSlider3D. Our technique consists of three stages: (1) generating image conditions from the 3D object for both reconstruction and attribute editing, (2) steering the generation process to produce 3D variations at different attribute intensities, and (3) mapping variations to perceptually consistent slider positions with adaptive bounds. We first describe the technical foundation of our technique, then detail each stage.


\subsection{Technical Foundation}
\label{sec:techfoundation}
SemanticSlider3D builds on TRELLIS \cite{xiang2025structured}, a state-of-the-art 3D generation model that represents 3D objects using Structured LATents (SLAT)---local latent vectors defined on a sparse 3D voxel grid that jointly encode both geometry and appearance. These structured latents can be decoded into multiple output formats, including 3D Gaussians, Radiance Fields, and meshes. We selected TRELLIS after evaluating several 3D generation models for two reasons: first, its flow matching-based pipeline supports inference-time intervention without per-attribute training, which scales better than LoRA-based approaches \cite{gandikota2024concept} when users need to define arbitrary attributes on the fly; second, its two-stage architecture ensures high-quality 3D generation.

TRELLIS generates 3D objects through a two-stage pipeline using flow matching models \cite{lipman2022flow}. The first stage generates the sparse structure (which voxels are active in a 3D voxel grid), and the second stage generates the latent vectors attached to these voxels, conditioned on text or image inputs. Multi-view renders of a 3D object ($\mathcal{O}$) can be used as image conditioning ($c_\mathcal{O}$) to generate the corresponding SLAT through the flow model's sampling process. This generation-based conversion is key to our editing approach, because the editing happens \textit{within} the sampling process itself. At each timestep, the flow model predicts a velocity $\textbf{v}_\theta(x_t, t, c)$ that guides the noisy sample $x_t$ toward a structured latent consistent with the conditioning $c$. 

While computing editing directions (vectors in the latent space that correspond to specific semantic changes) from contrastive conditions is well-established in 2D image sliders \cite{gandikota2024concept, ezra2025freesliders, ekin2026unreasonable}, applying this principle to a 3D structured latent space has been underexplored. We found that TRELLIS's flow matching formulation naturally supports training-free semantic steering: Given contrastive image conditions $c_+$ and $c_-$ corresponding to the positive and negative extremes of an attribute, a steering direction can be computed from the difference between their predicted velocities: $\textbf{d}_s = \textbf{v}_\theta(x_t, t, c_+) - \textbf{v}_\theta(x_t, t, c_-)$. By scaling $\textbf{d}_s$ with a factor $\alpha$ and adding it to the velocity predicted under the original object's conditioning $\textbf{v}_\theta(x_t, t, c_\mathcal{O})$, the sampling trajectory is steered toward different intensities of the target attribute.

We describe how each stage of our pipeline works in the following subsections.

\subsection{Generating Image Conditions}
We first render multiple camera views from the 3D object to get initial image condition, and then select and edit attribute-relevant views to construct contrastive image pairs for steering the generation in the latent space.

\subsubsection{Multi-View Rendering as Initial Condition}
\label{step1}
For a given 3D object, we sample $N$ cameras uniformly distributed across a sphere surrounding the object using the Hammersley quasi-random sequence \cite{wong1997sampling} \change{(Fig \ref{fig:pipeline}a)}. Each camera is placed at a fixed radius of 2.0 from the object's center with a 40\textdegree\ field of view, looking toward the center. We render a set of camera views $\{p_i\}_{i=1}^N$ using Blender following a similar process as TRELLIS. The rendered images are encoded as multi-view conditions $\{c_i\}_{i=1}^N$ via DINOv2 \cite{oquab2023dinov2}, which serve as the conditioning for reconstructing the original object through the flow model's sampling process.

\subsubsection{Generating Contrastive Editing Prompts}
For a user-specified continuous semantic attribute, we use an LLM (GPT-5.2) to generate a pair of editing prompts corresponding to the positive and negative extremes of the attribute, denoted as $(t^+, t^-)$ \change{(Fig \ref{fig:pipeline}b)}. 
Some attributes can be difficult to describe with a neutral statement (e.g., how traditional a sofa looks). In such cases, we also allow the user to provide only a prompt describing the change in the positive direction, and the LLM automatically generates the corresponding prompt pairs.
The LLM instruction prompt is provided in the Appendix \ref{appendix:prompt_generation}. These editing prompts are used to guide the generation of contrastive image pairs in the following steps.

\subsubsection{Selecting Edit-Relevant Views}
Rather than generating entirely new images from the editing prompts, which would introduce uncontrolled alteration beyond the target attribute, we edit the target attribute directly on the rendered camera views of the original object (\textbf{DG4}). This strategy ensures that the contrastive image pairs differ from the original only in the target attribute while preserving everything else.
However, not all camera views contain visual information relevant to the target attribute. For example, when editing the eye size of a 3D character, views from the back do not contain the relevant attribute for effective editing. To select edit-relevant views, we use the ImageReward score \cite{xu2023imagereward, jain2025adaptivesliders} to measure the alignment between the editing prompt and each camera view, where a higher score indicates greater relevance \change{(Fig \ref{fig:pipeline}c)}. However, irrelevant views can occasionally receive high scores due to visual ambiguity at certain camera angles. To mitigate this, we additionally use GPT-5.2 to verify whether the target attribute is actually present in each view (Appendix \ref{appendix:view_relevance}). Using this two-step pipeline, we select the $M$ most edit-relevant views $\{\hat{p}_i\}_{i=1}^M$ for constructing contrastive image pairs.

\subsubsection{Generating Contrastive Image Pairs}
\label{sec:edit views}
Using GPT Image 1.5, we edit each of the $M$ selected views with both $t^+$ and $t^-$, appending a reassurance instruction that explicitly prevents modification of attributes irrelevant to the target edit (\textbf{DG4}). Because the contrastive image pairs are used to compute a steering direction through velocity differences rather than to directly reconstruct a 3D object, strict multi-view consistency across the edited images is not required \cite{gandikota2024concept, ezra2025freesliders}. This produces contrastive image pairs $\{(\hat{p}_i^+, \hat{p}_i^-)\}_{i=1}^M$ that differ primarily in the target attribute while preserving other visual characteristics of the original view \change{(Fig \ref{fig:pipeline}d)}. The image pairs are then encoded as condition pairs $\{(\hat{c}_i^+, \hat{c}_i^-)\}_{i=1}^M$ via DINOv2.

\subsection{Editing via Generation}

As described in \S~\ref{sec:techfoundation}, semantic editing can be achieved by computing a steering direction from contrastive conditions and injecting it into the sampling process. Since our contrastive conditions span $M$ view pairs, we compute the steering direction at each timestep $t$ as the average velocity difference across all contrastive pairs:
\begin{equation}
\mathbf{d}_s(x_t, t) = \frac{1}{M}\sum\nolimits_{i=1}^{M} \left[\mathbf{v}_\theta(x_t, t, \hat{c}_i^+) - \mathbf{v}_\theta(x_t, t, \hat{c}_i^-)\right]
\end{equation} 
Averaging across view pairs produces a more robust editing direction that captures the target attribute from multiple perspectives while canceling view-specific noise, \change{isolating the edit to the target attribute (\textbf{DG4}).}

\change{To improve reconstruction quality (\textbf{DG1}) and further preserve unintended attributes (\textbf{DG4}), we apply classifier-free guidance \cite{ho2022classifier} to the average velocity predicted under the original object's multi-view conditioning:
\begin{equation}
\mathbf{\bar{v}}(x_t,t) = \mathbf{v}_\theta(x_t,t,\varnothing) + w \left[ \frac{1}{N}\sum\nolimits_{i=1}^{N}\mathbf{v}_\theta(x_t,t,c_i) - \mathbf{v}_\theta(x_t,t,\varnothing) \right]
\end{equation} 
where $\varnothing$ is the empty condition and $w$ the guidance strength.}

The edited velocity at timestep $t$ is then computed by adding the scaled steering direction to \change{this guided velocity:
\begin{equation}
\mathbf{v}_{\text{edit}}(x_t, t) = \mathbf{\bar{v}}(x_t, t) + \alpha \mathbf{d}_s(x_t, t)
\end{equation}}
where $\alpha$ controls the edit intensity: positive values steer towards the positive extreme of the attribute and negative values towards the negative extreme, enabling bidirectional editing along the semantic axis. \change{$\mathbf{\bar{v}}(x_t, t)$} reconstructs the original object from its multi-view conditions, while \change{$\alpha \mathbf{d}_s(x_t, t)$} steers the sampling trajectory toward the desired edit.

This steering is applied to both stages of TRELLIS's pipeline, allowing the edit to affect both coarse geometry and fine appearance (\textbf{DG1}) \change{(Fig \ref{fig:pipeline}e)}.

\subsection{Mapping Latent Variations to the Slider}
By adjusting the scaling factor $\alpha$, we can produce different degrees of variation along the semantic attribute in the latent space. However, to build a usable slider-based editing experience, we need to translate these latent-space variations into visually meaningful and consistently spaced slider positions (\textbf{DG2}, \textbf{DG3}).

\subsubsection{Determining Slider Bounds}
\label{sec:bounds}
To maximize the range of variation while ensuring that both ends of the slider maintain sufficient generation quality, we adopt an adaptive bound search inspired by prior work on 2D sliders \cite{jain2025adaptivesliders}.

We first set empirical outer bounds \change{$\pm B$ for $\alpha$}, determined via preliminary experiments to be large enough to cover the effective editing range for most object-attribute pairs, but not so large that the search wastes time evaluating clearly invalid generations (e.g., broken meshes). Starting from $\alpha = B$ and $\alpha = -B$, we incrementally decrease $|\alpha|$ inward from both ends with a step size $s$. At each candidate $\alpha$, we generate the corresponding 3D variation and evaluate it using a VLM (GPT-5.2), which checks for the presence of structural degradation or drastic appearance changes irrelevant to the target attribute across four views rendered at 90\textdegree\ horizontal intervals with 30\textdegree\ elevation (referred to as \textit{key-frames} hereafter) \change{(see Appendix \ref{appendix:quality_check} for instruction prompt)}. The search stops in each direction at the first $\alpha$ that passes this quality check. This yields an adaptive range $[\alpha_{\min}, \alpha_{\max}]$ tailored to each object-attribute pair (\textbf{DG2}).

\subsubsection{Perceptual Distance Estimation}
\label{sec:perceptual}
Within the determined bounds $[\alpha_{\min}, \alpha_{\max}]$, we sample $K$ evenly spaced values of $\alpha$ and generate a 3D variation for each \change{(Fig \ref{fig:mapping}a)}. To map these variations to perceptually meaningful slider positions, we measure the visual distance between each variation and the original object using LPIPS \cite{zhang2018perceptual}, a perceptual similarity metric closely aligned with human judgment, \change{widely used to evaluate perceptual fidelity in 3D reconstruction pipelines \cite{xiang2025structured, haque2023instruct, chen2026sam}.}

A challenge unique to 3D slider-based editing is that the generation process does not guarantee consistent object orientation across different $\alpha$ values, meaning two variations at different intensities may face different directions. \change{Comparing key-frames by rendering order would then measure differences between mismatched viewpoints, which would inflate perceptual distance measurements with orientation differences unrelated to the actual attribute change.
We found that the orientation offset between a variation and the original is almost always a multiple of 90\textdegree, so matching key-frames rendered every 90\textdegree~suffices.} We construct a $4 \times 4$ cost matrix of pairwise LPIPS distances between every key-frame (\S~\ref{sec:bounds}) of the original object and the variation, then solve a minimum-cost bipartite matching using the Hungarian algorithm \cite{kuhn1955hungarian} to \change{pair each original view with its closest-matching viewpoint in the variation.} The signed perceptual distance for each variation is then:
\begin{equation}
d(\alpha_k) = \frac{1}{4} \text{sgn}(\alpha_k) \sum\nolimits_{j=1}^{4} \text{LPIPS}(f_j^{\text{orig}}, f_{\sigma(j)}^{\alpha_k})
\end{equation}
where $f_j^{\text{orig}}$ is the $j$-th key-frame of the original object, $f_{\sigma(j)}^{\alpha_k}$ is the matched key-frame of the variation at $\alpha_k$, and $\sigma$ is the optimal assignment from the Hungarian algorithm, and $\text{sgn}(\alpha_k)$ assigns a positive or negative sign based on the editing direction.

\subsubsection{Slider Position Mapping \& Refinement}
\label{sec:mapping}
We adapt the perceptual distance-based slider mapping approach from AdaptiveSliders \cite{jain2025adaptivesliders} to our 3D setting.
We assign slider positions based on the signed perceptual distances: the most negative $d(\alpha_k)$ maps to $-100\%$ and the most positive to $+100\%$, with intermediate positions assigned proportionally \change{(Fig \ref{fig:mapping}b)}. This produces a mapping from $\alpha$ values to slider positions grounded in perceptual change rather than raw latent-space distance (\textbf{DG3}).

We then iteratively refine the slider by interpolating additional anchor points. Given a desired slider position (a percentage from -100\% to 100\% ), we interpolate the corresponding $\alpha$ from the existing mapping, generate the variation, compute its perceptual distance, and insert it as a new anchor to improve subsequent interpolations \change{(Fig \ref{fig:mapping}c)}. 
In practice, interpolated variations may not land precisely at the target percentage, as the latent space does not always support smooth linear interpolation of visual changes \cite{jain2025adaptivesliders, ezra2025freesliders}. Iterative refinement helps mitigate this.

Finally, we apply \change{the same VLM-based quality check as in \S~\ref{sec:bounds} (Appendix \ref{appendix:quality_check})} to filter out visually invalid variations within the bounds. Unlike 2D image generation, where variations within a reasonable $\alpha$ range are typically valid, 3D generation can produce occasional artifacts (e.g., broken geometry). Invalid variations are removed from the final slider. We then search for the nearest valid replacement by incrementally shifting the slider position by 1\% at a time until finding a variation that passes the quality check (\textbf{DG1}). \change{Fig \ref{fig:mapping}d shows a comparison before and after slider position mapping on a ``rustiness'' slider created for a 3D airplane.}
\section{Technical Validation}
To evaluate whether our technique satisfies the four design goals, we constructed a dataset of 50 object-attribute pairs spanning diverse 3D object categories and continuous semantic attributes, and conducted a \change{validation experiment with five human assessors.}

\subsection{Validation Dataset}
\subsubsection{Object Categories}
To ensure broad coverage, we selected 3D objects from categories commonly represented in established 3D object datasets \cite{objaverse, stojanov2021using, fu20213d}, including human characters, animals, plants, furniture, transportation, electronics, food, and toys.

\subsubsection{Attribute Categories}
Drawing on attribute types used in prior work on 2D image sliders \cite{jain2025adaptivesliders, gandikota2024concept, gandikota2025sliderspace}, we organized continuous attributes into two main categories: \textit{local attributes} and \textit{overall characteristics}.

\textit{Local attributes} target specific parts of the object, including (1) spatial properties such as eye size or the height of a chair backrest, and (2) facial expressions such as degree of smile.

\textit{Overall characteristics} affect the object as a whole, including (1) overall shape such as body shape, (2) style such as traditional vs.\ modern, (3) material properties such as surface roughness, (4) level of detail such as low-poly vs.\ photorealistic, and (5) temporal properties such as doneness of food or age of a person.

\subsubsection{Dataset Construction}
Two researchers collaboratively generated candidate object-attribute pairs from the above categories and reviewed each pair to ensure it represented a reasonable and editable continuous attribute for the given object. This process yielded 50 object-attribute pairs across 46 unique objects, with a few objects appearing in multiple pairs with different attributes. For each pair, we sourced 3D objects from CC-BY licensed assets in Objaverse \cite{objaverse} and from 3D-FUTURE \cite{fu20213d}. We queried each dataset using keywords derived from the target object name based on available metadata, randomly retrieved 20 candidate objects per query, and manually selected the one that best matched the intended object with sufficient geometric and appearance quality. 

\subsection{Apparatus}
To assess how well SemanticSlider3D satisfies the four design goals, we compared it with a baseline as a reference. Since no existing method directly supports slider-based continuous semantic editing of 3D objects, we constructed a baseline by combining Concept Sliders \cite{gandikota2024concept}, a state-of-the-art 2D image slider, with SAM 3D \cite{chen2026sam}, a state-of-the-art image-to-3D model, to represent the most straightforward off-the-shelf alternative: editing a 2D image along a slider and then lifting each variation to 3D independently. For both the baseline and SemanticSlider3D, we aimed to generate nine variations on the resulting slider, presenting the full spectrum of each target attribute.

\subsubsection{SemanticSlider3D Setup}
\label{val:ours}
For each object-attribute pair, we rendered $N=40$ camera views of the 3D object and selected $M=10$ most relevant views for constructing contrastive image pairs, balancing generation speed and quality. 
We set the \change{outer bound $B=5$ (sufficient to reach invalid generations for most object-attribute pairs), step size $s = 0.25$ (ensuring relatively fine granularity while keeping generation time reasonable), and guidance strength $w=3$ (balancing condition adherence with steerability).
In the slider mapping stage, we completed $R = 2$ rounds of refinement after sampling $K = 9$ initial $\alpha$ values.} The slider variations corresponding to nine evenly spaced percentages from $-100\%$ to $100\%$ were exported as GLB files. The entire pipeline was fully automatic and ran on a single A100 GPU.

\subsubsection{Baseline: Concept Sliders + SAM 3D}
Concept Sliders \cite{gandikota2024concept} train low-rank adaptors (LoRA) for each target attribute in a diffusion model's weight space, learning a parameter direction that corresponds to the desired semantic concept. At inference time, the trained adaptor is loaded and a single scalar weight controls the intensity of the edit. For each object-attribute pair, we trained a Concept Slider on SDXL-Turbo \cite{sauer2024adversarial}, a fast text-to-image diffusion model, using the same 10 contrastive image pairs (positive and negative) as in \S~\ref{val:ours}.

Since Concept Sliders operate on a single input image, we used GPT-5.2 to select the most representative and informative camera view from all 40 rendered views. To generate variations from this input image, we applied Regularized Newton Inversion (GRNI) \cite{samuel2023regularized} to convert the image into the diffusion model's latent space. Following their inversion setup \cite{samuel2023regularized}, we set the denoising strength to 0.6. We used a detailed caption of the 3D object as the inversion prompt, obtained using the captioning method described in \cite{xiang2025structured}.

We applied the same slider bound search and position mapping algorithm as SemanticSlider3D, with the outer bound set to $B=12$, larger than the $B=8$ used in prior work \cite{jain2025adaptivesliders}, to further ensure no valid variations at the extremes are missed. After finalizing the nine 2D slider variations, we segmented the object from the background using SAM 3 \cite{carion2025sam} with a basic object description as prompt, and lifted each segmented image to 3D using SAM 3D \cite{chen2026sam}. The resulting 3D objects were exported as GLB files. The entire baseline pipeline was fully automatic and ran on a single A100 GPU.

\subsubsection{Evaluation Interface}
We built a web-based evaluation interface (Fig ~\ref{fig:val_ui}) that presents human assessors with one object-attribute pair at a time. The top left panel displays the reference 3D object and the target attribute. The bottom panel shows two rows of nine 3D variations generated by SemanticSlider3D and the baseline, labeled Slider A and Slider B. The mapping between methods and labels is randomized for each object-attribute pair so that assessors cannot identify which method they are evaluating. Assessors can rotate all objects simultaneously to inspect them from different angles, adjust the cell window size for closer inspection.
The top right panel presents evaluation questions one at a time. For each object-attribute pair, assessors first rate Slider A on four criteria using a 7-point Likert scale: (1) range of variation, (2) quality of the 3D object, (3) variation consistency, and (4) preservation of irrelevant attributes. They then rate Slider B on the same four criteria. On each question, the meaning of the two ends on the scale is labeled. Finally, they indicate an overall preference between the two methods (Slider A, Slider B, or Can't Decide). In total, assessors answer nine questions per object-attribute pair before proceeding to the next.

\subsection{Procedure}
Five assessors participated in the technical validation, three from a software company and two from a local university. None of them had prior exposure to either method. Each assessor evaluated all 50 object-attribute pairs in a single session of 2 hours. Before beginning, assessors received an explanation of the task and four evaluation criteria (design goals) with image examples. To calibrate ratings, assessors first completed one practice trial, after which the researcher discussed their rationale and clarified any misunderstandings of the criteria. Assessors were also instructed to ignore minor color or lighting variations due to rendering conditions and monitor differences.

\subsection{Results}
\label{val:res}
We report results across the four evaluation criteria corresponding to our design goals, as well as overall preference, \change{breaking results down by the seven attribute categories. For each criterion, we report mean ratings for both methods across all 50 object-attribute pairs over the five assessors. 
Because a two-tailed Wilcoxon signed-rank test cannot reach significance with five assessors ($p \geq .063$ by construction), we report descriptive statistics and inter-rater reliability (IRR). To assess IRR, we used Gwet's AC$_1$ for the overall preference and AC$_2$ for the ordinal Likert-scale ratings, both robust to skewed rating distributions \cite{gwet2014handbook, tjia2025novel}.}
For object-attribute pairs where a method received a rating of 2 or below for range of variation, we excluded that method's variation consistency rating from analysis, as a near-zero range trivially implies smooth variation, making the consistency question meaningless. Fig \ref{fig:likert-val} details the Likert responses from the five assessors, \change{and mean ratings are also summarized as a radar plot (Fig \ref{fig:radar}).}

\subsubsection{Quality of the 3D Object (DG1)}
Overall, SemanticSlider3D received high ratings ($Mean=5.70$, $SD=0.38$) \change{with almost perfect agreement across assessors ($AC_2 = 0.96$), indicating that SemanticSlider3D produces high quality 3D Objects. The baseline was also rated positively ($Mean=4.66$, $SD=0.75$), with substantial agreement ($AC_2 = 0.72$).} Our method scored highest on temporal ($Mean=5.98$, $SD=0.28$) and style attributes ($Mean=5.93$, $SD=0.43$), and lowest on local spatial attributes ($Mean=5.18$, $SD=0.47$), where fine-grained geometric edits occasionally introduced minor artifacts.
Material attributes were rated towards the higher end for both methods (Ours: $Mean=5.62$, $SD=0.41$; Baseline: $Mean=5.47$, $SD=0.48$).

\subsubsection{Range of Variation (DG2)}
SemanticSlider3D received high ratings indicating wide range of variation ($Mean=6.60$, $SD=0.20$) \change{with very high IRR ($AC_2 = 0.86$),} and across all categories, ranging from 6.36 ($SD=0.36$) on local spatial attributes to 6.80 ($SD=0.17$) on temporal attributes. Local spatial was the lowest-scoring category for our method, possibly because localized geometric changes (e.g., eye size, backrest height) are harder to steer in the latent space. Baseline ratings were on the lower end of the scale ($Mean=2.87$, $SD=0.32$), \change{though with low agreement ($AC_2 = 0.39$).}

\subsubsection{Variation Consistency (DG3)}
SemanticSlider3D showed high variation consistency ($Mean = 5.69, SD = 0.25$) \change{and almost perfect agreement among assessors ($AC_2 = 0.85$).
While the baseline's mean was also high (Mean=5.22, SD=1.07), ratings varied widely across assessors ($AC_2 = 0.54$).}
Our method scored consistently high on material ($Mean = 6.29, SD = 0.32$) and temporal attributes ($Mean = 6.20, SD = 0.33$). Assessors were moderately satisfied with overall shape ($Mean = 4.36, SD = 0.23$). This may be because shape edits require coordinated geometric changes across many parts of the object, making it difficult to produce evenly spaced variations compared to appearance-dominant attributes like style or material, where the overall structure can remain stable while the surface properties change.

\subsubsection{Preservation of Irrelevant Attributes (DG4)}
SemanticSlider3D demonstrated high preservation of irrelevant attributes overall ($Mean=5.69$, $SD=0.43$), \change{with high IRR ($AC_2 = 0.77$). The baseline received moderate ratings on this criterion ($Mean = 4.20, SD = 0.55$) with low IRR ($AC_2 = 0.39$).} Our method scored high on temporal ($Mean=6.20$, $SD=0.41$) and style attributes ($Mean=6.20$, $SD=0.58$), and \change{lower} on local spatial attributes ($Mean=4.82$, $SD=0.80$). The baseline showed low performance on facial expressions ($Mean = 2.52, SD = 1.00$). This is because each step of the pipeline—image inversion, object segmentation, and 2D-to-3D conversion—can introduce fidelity loss, leading to unintended changes beyond the target expression. Local spatial attributes were the \change{lower-scoring category for both methods} (Baseline: $Mean=4.24$, $SD=0.69$), where both methods found it relatively challenging to fully isolate the edit from irrelevant features.

\subsubsection{Overall Preference}
\change{All five assessors preferred SemanticSlider3D over the baseline with substantial agreement ($AC_1 = 0.79$), with individual preference rates ranging from 72\% to 96\%, consistently across categories.}

\subsubsection{Summary}

\change{SemanticSlider3D's absolute ratings were above 5.0 on all four criteria, suggesting that our method satisfies the four design goals as perceived by human assessors.}
Across attribute categories, our method performed strongest on temporal and style attributes, and was more limited on spatial attributes (both local and overall), where precise geometric changes are harder to steer smoothly in the latent space.
\section{User Study}
We conducted an exploratory user study with six participants who had prior experience with 3D tools. We aimed to understand how users would use SemanticSlider3D, and get their feedback on its usability and utility for 3D prototyping. 

\subsection{Apparatus}
\label{apparatus}
To this end, we built an interactive playground as a technology probe \cite{hutchinson2003technology} using React \cite{react} and Flask \cite{flask-sio}.
\change{Users could generate 3D objects from text (TRELLIS text-to-3D) or image (SAM 3D), and create sliders using SemanticSlider3D. We used the same SemanticSlider3D setup as detailed in \S~\ref{val:ours}, except with $K=4$ (user-adjustable) initial samples and $R=1$ refinement round to balance generation time and granularity.}
The interface (Fig \ref{fig:user}) consists of the following components:

\begin{figure*}[h]
  \includegraphics[width=\textwidth]{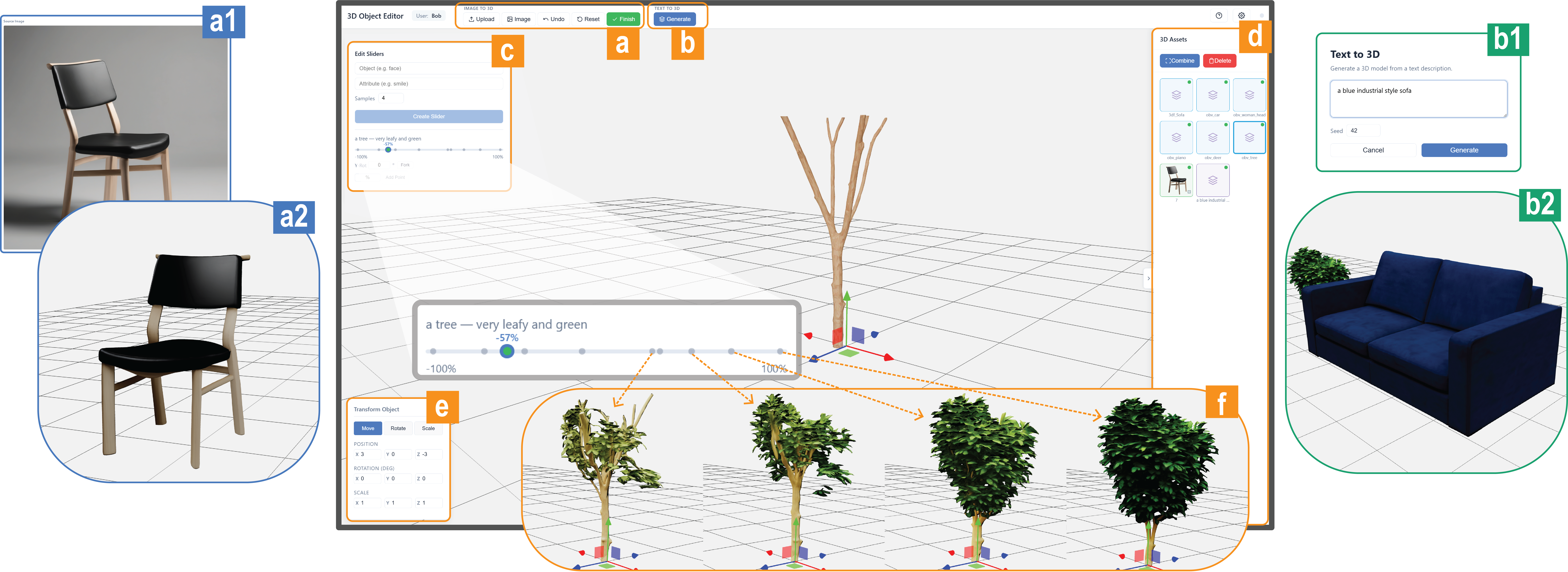}
  \vspace{-3ex}
  \caption{\change{The interactive playground interface used in our study. Users generate 3D objects via \textbf{image-to-3D} (a: upload and segment a reference image (a1) to produce a 3D object (a2)) or \textbf{text-to-3D} (b: type a description (b1) to generate a 3D object (b2)). The \textbf{Slider Panel} (c) allows users to specify an object and semantic attributes, create a slider, and add new anchor points at desired percentages; as the user moves the slider handle on created sliders, the corresponding variation is displayed in-place in the viewport (f). The \textbf{Sidebar} (d) displays thumbnails of all objects in the scene. The \textbf{Transform Panel} (e) provides controls for positioning, rotating, and scaling selected objects.}}
  \label{fig:user}
  \vspace{-3ex}
\end{figure*}
 
\begin{itemize}[leftmargin=1em]
    \item \textbf{3D Viewport}: The central workspace where users can view and manipulate 3D objects. Objects can be selected, repositioned, rotated, and scaled using on-screen gizmos or the \textbf{Transform Panel (e)}.
    \item \textbf{Image-to-3D (a)}: Users can upload a reference image (a1), segment a target object by clicking on it (via SAM 2 \cite{ravi2025sam}), and lift the segmented object into a 3D object (a2).
    \item \textbf{Text-to-3D (b)}: Users can type a text description (b1) to generate a 3D object (b2).
    \item \textbf{Slider Panel (c)}: Users specify an object and a semantic attribute in text fields, then click ``Create Slider'' to generate the corresponding semantic slider with initial anchors points. Before generating, users can adjust how many anchor points to sample --- more points produce a smoother slider but take longer to generate. The created slider displays variations along the semantic attribute, and users can move the slider handle to preview \textbf{different variations (f)}. An ``Add Point'' feature allows users to request a new anchor point at a specified percentage. The system interpolates the corresponding variation from the existing slider mapping and generates it, progressively filling in the slider for finer exploration.
    \item \textbf{Sidebar (d)}: Thumbnails of all 3D objects in the scene. Clicking a thumbnail selects the corresponding object in the viewport. Right-clicking provides options to hide/unhide.
\end{itemize}

\change{Our playground was served on two A100 GPUs (four processes each), allowing at most two sliders to be built concurrently. 
Across all sliders built during the study sessions (\S~\ref{procedure}), build time under default setting ($K=4$, $R=1$) averaged 726.3s ($SD = 183.2$). It varied across object-attribute pairs because failed quality checks required additional generations and mesh complexity affected per-variation generation time (43s at $\sim$1.2k triangles to 84s at $\sim$50k triangles). Table~\ref{tab:timing} reports the full per-stage time breakdown for our 3D editing pipeline.}

\subsection{Participants}
We recruited six participants (two male, four female, ages 22--31, $Mean=26$, $SD=3.16$) with prior exposure to 3D prototyping from two universities. Four were graduate students in computer science and two in mechanical engineering. P6 also had a background as a sculpture and animation artist. Participants had varying levels of 3D prototyping experience, ranging from 1 month to 10 years, using tools such as Autodesk Fusion, SolidWorks, Blender. In addition, P1, P2 and P3 had prior exposure to online 3D generation services such as Meshy AI \footnote{https://www.meshy.ai/}. Each participant received a \$25/hr compensation via Amazon gift cards. Participants' demographic information and 3D prototyping experience are detailed in Table \ref{tab:participants}

\subsection{Study Design}
\label{procedure}
\subsubsection{Procedure}
We conducted in-person study sessions, each lasting about 150 minutes. The study consisted of three phases.

In the \textit{introduction phase}, the researcher explained the purpose of the study and walked participants through the interface, which was preloaded with six objects and sliders (e.g., a car with a headlight size slider, a deer with a realisticness slider) to familiarize participants with the range of attributes the system supports. The walkthrough covered how to create sliders by specifying an object and a semantic attribute, preview variations by moving the slider, and add new anchor points. Participants then had 5 minutes to freely explore before proceeding.

In the \textit{task phase}, participants completed two tasks. In \textit{Task 1}, participants were asked to think of a 3D prototyping scenario with a specific goal in mind, then create two 3D objects (via text-to-3D or image-to-3D) and a slider for each. In \textit{Task 2}, participants created two additional sliders based on existing objects in the scene, either by selecting the original object or by copying an anchor point variation as a new starting point. \change{Once built, participants could browse variations without latency; adding a new anchor point took about 90s.} Participants were asked to think aloud throughout. After each slider creation, the researcher asked about their subjective experience, including whether results aligned with expectations and their next steps toward their goal. During generation wait times, the researcher collected demographic information and participants' prior experience with 3D prototyping and generative AI tools.

In the \textit{exit interview}, participants responded to questions adapted from the Creativity Support Index~\cite{cherry2014quantifying} and slider-specific questions on the usefulness of different slider components. \change{All items were phrased as agreement statements (e.g., ``The system helped me fully explore the space of potential designs'') and rated on a 7-point Likert scale from 1 (strongly disagree) to 7 (strongly agree).} We also collected general feedback about their strategies during 3D prototyping with SemanticSlider3D and suggestions for improvement.

\subsubsection{Data Collection and Analysis}
We transcribed all recordings and analyzed the transcripts using thematic analysis~\cite{braun2006using}. The first author coded all transcripts using open coding, and then iteratively generated and refined themes while reviewing participants' artifacts. The resulting codes and themes were reviewed and discussed with the research team.

\subsection{Findings}
\label{findings}
We detailed participants' individual 3D prototyping tasks and goals in Appendix \ref{participant tasks}. In this section, we report participants' experience with our technology probe via qualitative feedback and Likert-scale responses.
 
\subsubsection{Slider Variations for Exploration}
\begin{figure}[h]
  \includegraphics[width=0.47\textwidth]{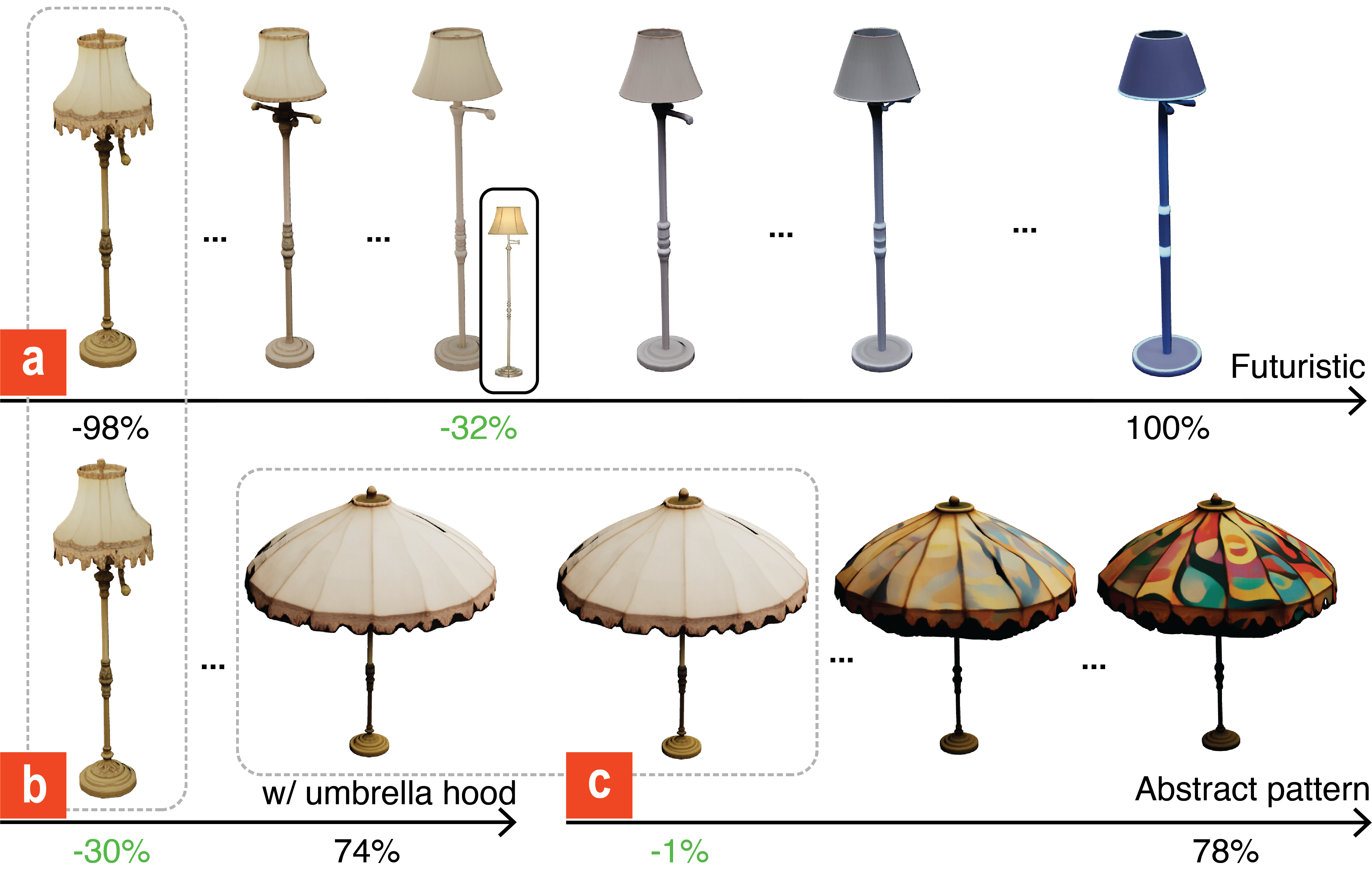}
  \vspace{-3ex}
  \caption{\change{P3 created a (a) ``futuristic style'' slider for a floor lamb generated from an image, and chose the -98\% variation to further refine with a (b) ``umbrella-shaped hood'' slider and chose 74\% variation to further edit with a (c) ``pattern abstractness'' slider to get a desired pattern. Green \% = input object's slider position (same in later figures).}}
  \label{fig:p3}
  \vspace{-2ex}
\end{figure}

\begin{figure}[h]
  \includegraphics[width=0.47\textwidth]{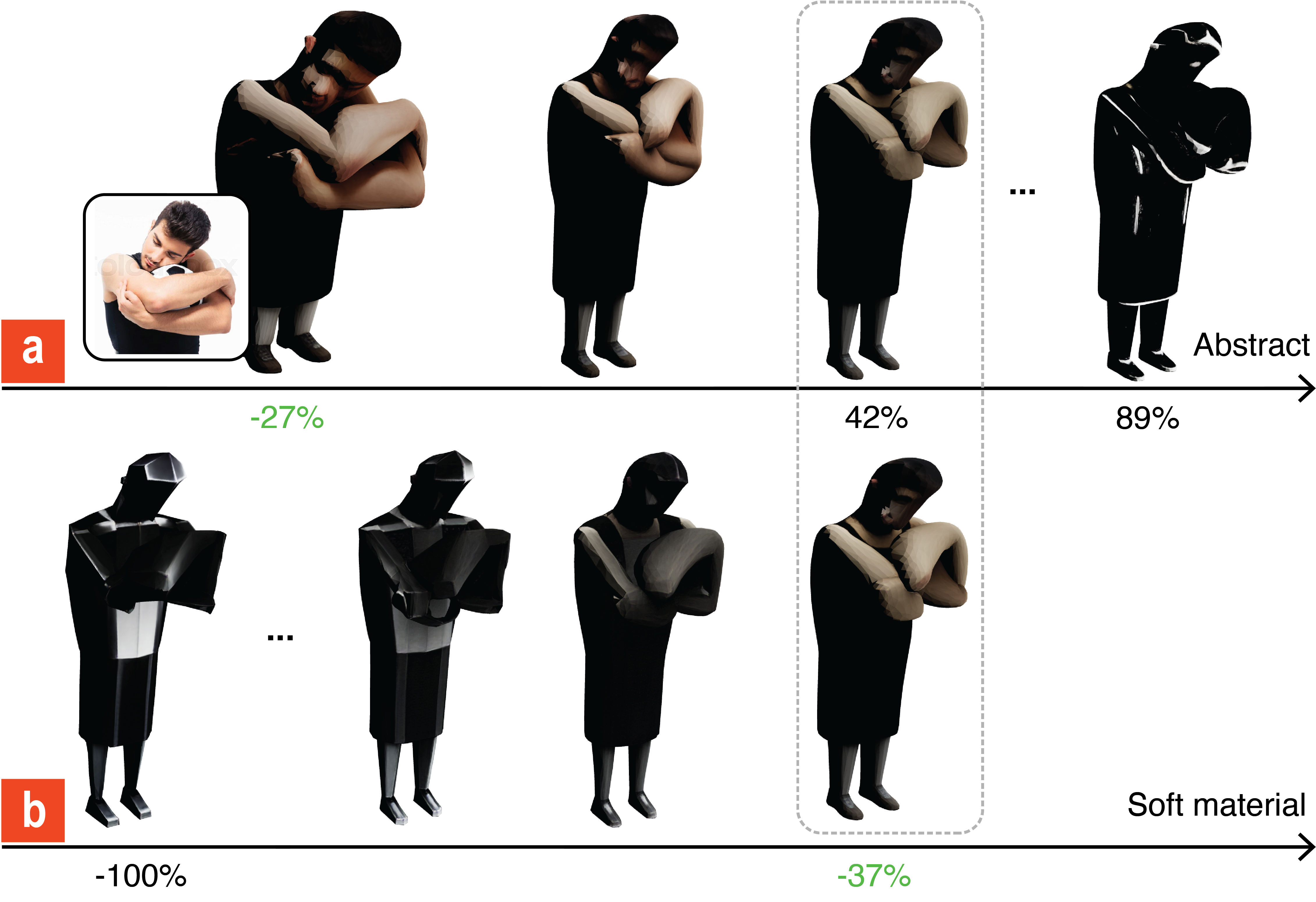}
  \vspace{-3ex}
  \caption{\change{P6 created an (a) ``abstractness'' slider on an object generated from an image (a man hugging a ball) sourced online, with the goal of designing a sculpture with the same gesture. She then chose a variation at 42\% and further explored material use with a (b) ``hard vs. soft'' slider where she found a good design at -100\%.}}
  \label{fig:p6}
  \vspace{-2ex}
\end{figure}

Participants broadly appreciated the slider's ability to generate diverse design variations along a semantic attribute, all from a single object and prompt specification (P1, P2, P3, P6). The slider was rated favorably for facilitating diverse exploration ($Mean = 6.17$, $SD = 1.17$) and for supporting decisions on what other design variations to try ($Mean = 6.17$, $SD = 0.75$). Participants also found that \textbf{slider variations can bring unexpected inspiration} to their prototyping (P1, P2, P3, P6). P3 originally wanted a very modern and futuristic floor lamp, but ended up being attracted by a retro variation (Fig \ref{fig:p3}): \textit{``because it (retro style lamp) has much more features or much more designs, instead of just a very simple hood (a modern variation).''} P2 additionally commented that \textbf{existing slider variations could trigger new exploration directions}, \change{while brainstorming the design of a prosthetic leg,} \textit{``I was already just thinking about the style... but after I see different options provided by the (futuristic style) slider and now I think `oh there could be more things added to it,' so now it inspired me to think about the length and authenticity but the slider gave the trigger.''} Despite these merits, two participants (P3, P6) wanted to explore beyond the current 100\% endpoint. For example, P6 would like to see more abstract versions of her sculpture (Fig \ref{fig:p6}) beyond the allowed range, suggesting that users should be granted more authority over what is considered out-of-range for attributes like artistic styles.

Participants also commented on the \textbf{granularity} of the slider's variations. P4 valued the slider's ability to show intermediate states that would be difficult to obtain through discrete prompting with current GenAI tools, such as seeing the full spectrum of color as an apple progresses through ripeness. However, three participants (P1, P3, P5) noted that the \textbf{percentage placement of anchor points was sometimes unpredictable}. This is likely due to the nonlinearity of the latent space (\S~\ref{sec:mapping}). P1 commented, \textit{``specifying where on the slider that (the new point) ends up going is difficult.''} These observations suggest that while the slider was effective at producing diverse variations, the perceived continuity did not always meet expectations.

\subsubsection{\change{Expressing Design Intents via Sliders}}
While the slider facilitated diverse exploration, participants also reflected on how the text-based attribute specification shaped their ability to express design intent. Overall, participants reported being able to be expressive ($Mean = 5.67$, $SD = 1.03$). 
\change{We observed that for vague attributes (e.g., "futuristic looking"), the LLM converged on a single concrete interpretation, whereas for specific attributes, the generated prompts closely followed the user's wording. In our study, four participants (P1, P4-P6) moved from broad concepts in Task 1 to more specific attributes in Task 2, producing outputs better matching their intent.}

\begin{figure}[h]
  \includegraphics[width=0.47\textwidth]{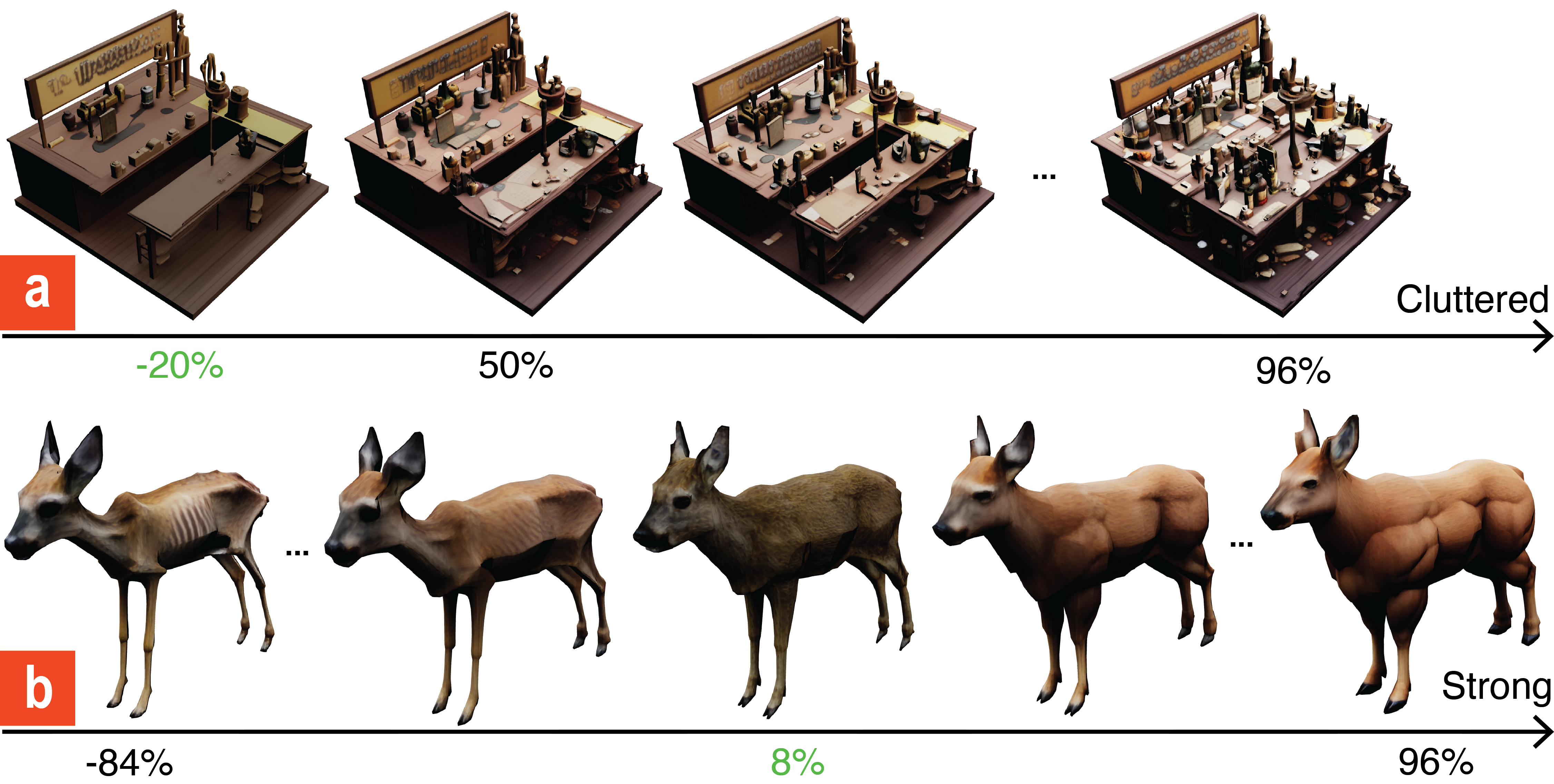}
  \vspace{-3ex}
  \caption{\change{(a) P1: no entrance for bartender at 96\% on a ``clutteredness'' slider for a bar counter; (b) P5: skinny deer at -84\% on a ``strong'' slider.}}
  \label{fig:p51}
  \vspace{-2ex}
\end{figure}

Seeing slider results helped \textbf{concretize and clarify participants' mental images} of the attributes they wished to edit (P1, P2, P4, P5). For example, P5 simply typed a single word ``strong'' in the attribute text box for a deer object, hoping to see a range of body shapes without a concrete idea of what she was looking for. However, after viewing the full spectrum, especially seeing the very skinny deer on the negative side (Fig \ref{fig:p51}b), she realized the attribute she actually wanted to adjust was \change{maturity (age)} rather than muscle mass. This suggests that the slider serves not only as an editing tool but also as a mechanism for users to calibrate their own understanding of their design intent.

Beyond this reflective use, participants also identified opportunities to enrich the attribute specification process. P1 and P5 wanted \textbf{system-provided guidance on prompt refinement}, and P2 wanted \textbf{attribute suggestions} given a 3D object, similar to features in existing creativity support tools~\cite{liu20233dall, jain2025adaptivesliders}. P2 and P6 suggested \textbf{using images to define attributes} as an alternative to text, because this is more natural for ``visual people.'' Three participants (P1, P3, P6) further expressed the \textbf{need for defining intermediate anchor points to allow mixing concepts}, similar to PromptPaint's prompt mixing \cite{chung2023promptpaint}. P6 envisioned a material slider from water to stone: \textit{``let's say water to stone. Maybe then the middle can be ocean or shoreline, like where the water and land meet... And then the sculpture can be something like... half water and... half sand... that's like in between.''}

\subsubsection{Integration into Current Workflow}
In general, participants were satisfied with the artifact they created: P5 and P2 wanted to use outputs directly for 3D printing and adding variations to a VR asset library, respectively; P3 and P6 began considering physical manufacturing processes; and P1 indicated his prototypes were ready for mesh-level refinement in 3D authoring tools.

Participants discussed how the slider-based technique could fit within their existing 3D prototyping practices. All participants considered the slider useful for rapid 3D prototyping ($Mean = 6.33$, $SD = 1.21$) and as a valuable addition to their current workflows ($Mean = 6.00$, $SD = 1.26$).

P3, an experienced mechanical engineer, explained how our technique can support rapid prototyping: \textit{``with one prompt, I was able to get a slider that have multiple designs on that. But if I do text-to-image, I would have to do multiple prompting to get different type of images then multiple prompts to convert that image to 3D. So it's a lengthy process... The reason we don't do 3D-initial 3D prototyping is because the time it takes to [generate] prototype in 3D. So this slider-based tool reduce that time by a lot.''} Beyond speed, P1 appreciated that the slider could \textbf{quantify subjective attributes by percentages}, giving designers a more concrete estimate of the amount of edit before reaching their goal, which is beneficial to the prototyping process regardless of whether slider-based editing is used downstream.
 
While promising for improving current workflows, some participants highlighted \textbf{practical considerations around functional feasibility}. For example, P1 noticed that some bar counter variations lacked an entrance for bartenders (Fig \ref{fig:p51}a), requiring further manual adjustment. 
However, P3 argued that such structural constraints are better addressed with conventional CAD tools, while the slider should focus on discovering visual attribute variations. This complementary view was echoed by participants' experiences with spatial attributes (P2, P5): both noted quality degradation for fine-grained geometric edits (e.g., gaze direction), but considered this acceptable given that conventional 3D authoring tools handle such adjustments more reliably.

\section{Discussion}
In this section, we reflect on findings from both the technical validation and the user study. We discuss the unique challenges of 3D continuous editing, how the slider interaction shapes human-AI co-creation during 3D prototyping, practical tradeoffs between latency and slider granularity, and tensions between system-imposed constraints and user autonomy.

\subsection{3D Editing via 2D vs.\ Direct Latent Steering}
Simply applying existing 2D image sliders and lifting the results to 3D seems like an intuitive solution, but our technical validation suggests that operating directly in the 3D latent space, as SemanticSlider3D does, is more effective. The baseline (Concept Sliders + SAM 3D) consistently struggled to produce meaningful variation, and its multi-step pipeline (image inversion, 2D editing, segmentation, 2D-to-3D conversion) introduced compounding fidelity loss that degraded both quality and attribute preservation, aligning with prior findings that 2D-assisted 3D generation often yields lower quality due to multi-view inconsistency \cite{xiang2025structured, shi2023mvdream}.

By contrast, SemanticSlider3D steers the generation directly in the 3D latent space, where geometry and appearance are jointly represented, avoiding the information loss of 2D-3D projection and naturally preserving multi-view consistency.
\change{However, spatial attributes (local and overall) remain challenging, as they require precise geometric changes that are hard to control through latent steering. A user-specified 3D structure prior indicating the desired geometric change at each slider point, inspired by \cite{mueller2025geodiffusion}, could potentially enable more controllable geometric edits.
More broadly, our latent steering approach could generalize to other 3D generation pipelines that support multi-view contrastive conditioning and a flow or diffusion sampler.}

\subsection{Slider-Based Human-AI Co-Creation}
Our exploratory user study uncovered the potential of SemanticSlider3D for human-AI co-creation during the prototyping process. Participants reported that slider variations brought unexpected inspiration and triggered new exploration directions that they had not initially considered. This finding echoes prior work on how generative AI can expand the design space by surfacing possibilities beyond what users would envision on their own \cite{liu20233dall, duan2025designfromx}.

However, the current interaction between the slider and the user is largely one-directional: the system generates variations and the user evaluates them. As noted in our findings, anchor point placement did not always match participants' expectations, primarily due to the nonlinearity or sparsity in the latent space which can't be fixed perfectly using LPIPS normalization techniques alone if the extent of the problem is too large. One promising direction to address this is to allow users to directly modify 3D variations on the slider---for instance, manually adjusting a variation to indicate what a desired percentage on the semantic axis should look like. This would turn the slider into a bidirectional interaction, where the user's manual edits inform the system's understanding of the attribute and help calibrate subsequent generations. Such a human-in-the-loop approach could help the user and the model converge on a shared understanding of the editing direction more efficiently, echoing prior work \cite{evirgen2022ganzilla}.

\subsection{Latency and Slider Granularity}
A practical tension in SemanticSlider3D is between slider granularity and generation time. Each slider took approximately 12 minutes to generate initial anchors, \change{and each additional anchor point requires another 90s (Table~\ref{tab:timing}).} Although smaller than the pretraining requirements for existing 2D LoRA based approaches, this latency limited the number of anchor points we could provide during the study, which could affect the perceived smoothness of the slider. 

This tradeoff between granularity and latency is not unique to our work. Dreamcrafter addressed a similar challenge by introducing 2D proxy representations for high-latency 3D generation and editing \cite{vachha2025dreamcrafter}. A similar strategy could benefit SemanticSlider3D: lightweight 2D previews could give users immediate feedback on the editing direction while the full 3D variations generate in the background. 
\change{Additionally, directly viewing Gaussian Splats could reduce latency by skipping mesh conversion (exporting to GLB), and running on more resource-intensive compute will make generation faster, enabling denser sampling and smoother sliders.}
In the meantime, future work could explore progressive refinement strategies that prioritize generating anchor points in regions where the user is actively exploring.

\subsection{Transparency and Freedom of Control}
Our study surfaced tensions between system automation and user awareness. Under the hood, SemanticSlider3D augments the user's attribute prompt via AI models to generate contrastive editing prompts, selects edit-relevant views, and determines slider bounds---all without user visibility. While this automation reduces user burdens, it also means the system is making interpretive decisions that can silently deviate from the user's original intent. 
Making these intermediate steps transparent, such as showing users the generated contrastive prompts and the selected views before generating the slider, could help users verify that the system's interpretation matches their intent and intervene early when it does not.

Beyond transparency, participants expressed a desire for greater freedom of control. As reported in our findings, participants wanted to explore beyond predetermined bounds (P3, P6), use images to define attributes (P2, P6), and mix concepts at intermediate anchor points (P1, P3, P6). These suggestions all indicate a need for more control over the editing process. This would allow users to calibrate the tool to their own creative standards and context while still benefiting from the system's default setting as a starting point.

\subsection{Limitations \change{\& Future Work}}
Our work has limitations which should be looked at in future work. First, the VLM-based quality checks (GPT-5.2) are not always reliable. Sometimes structurally valid variations were rejected, while variations with visible artifacts occasionally passed. Allowing users to override or review flagged variations could help mitigate this. Future work should also investigate more reliable quality assessment methods by incorporating 3D structural information rather than relying solely on 2D rendered views. Second, our validation dataset focused on common everyday objects with limited coverage of artistic or abstract attributes. Future work should expand the dataset to better evaluate the technique's applicability in creative and expressive contexts. We will open source our dataset to push this work forward. \change{Lastly, our technical validation involved a small number of assessors. A larger pool would enable statistically significant comparisons with the baseline.}

\section{Conclusion}
We presented SemanticSlider3D, a training-free technique for continuous, slider-based semantic editing of 3D objects. Our technical validation showed that SemanticSlider3D produces diverse, consistent, and high quality variations without altering irrelevant attributes, and was overwhelmingly preferred over a baseline. Our exploratory user study further showed that the technique facilitated exploration of design alternatives and was perceived as a useful addition to 3D prototyping workflows.

\bibliographystyle{ACM-Reference-Format}
\bibliography{reference}


\begin{thebibliography}{79}


\ifx \showCODEN    \undefined \def \showCODEN     #1{\unskip}     \fi
\ifx \showISBNx    \undefined \def \showISBNx     #1{\unskip}     \fi
\ifx \showISBNxiii \undefined \def \showISBNxiii  #1{\unskip}     \fi
\ifx \showISSN     \undefined \def \showISSN      #1{\unskip}     \fi
\ifx \showLCCN     \undefined \def \showLCCN      #1{\unskip}     \fi
\ifx \shownote     \undefined \def \shownote      #1{#1}          \fi
\ifx \showarticletitle \undefined \def \showarticletitle #1{#1}   \fi
\ifx \showURL      \undefined \def \showURL       {\relax}        \fi
\providecommand\bibfield[2]{#2}
\providecommand\bibinfo[2]{#2}
\providecommand\natexlab[1]{#1}
\providecommand\showeprint[2][]{arXiv:#2}

\bibitem[Achlioptas et~al\mbox{.}(2023)]%
        {achlioptas2023shapetalk}
\bibfield{author}{\bibinfo{person}{Panos Achlioptas}, \bibinfo{person}{Ian Huang}, \bibinfo{person}{Minhyuk Sung}, \bibinfo{person}{Sergey Tulyakov}, {and} \bibinfo{person}{Leonidas Guibas}.} \bibinfo{year}{2023}\natexlab{}.
\newblock \showarticletitle{ShapeTalk: A language dataset and framework for 3d shape edits and deformations}. In \bibinfo{booktitle}{\emph{Proceedings of the IEEE/CVF conference on computer vision and pattern recognition}}. \bibinfo{pages}{12685--12694}.
\newblock


\bibitem[Barda et~al\mbox{.}(2025)]%
        {barda2025instant3dit}
\bibfield{author}{\bibinfo{person}{Amir Barda}, \bibinfo{person}{Matheus Gadelha}, \bibinfo{person}{Vladimir~G Kim}, \bibinfo{person}{Noam Aigerman}, \bibinfo{person}{Amit~H Bermano}, {and} \bibinfo{person}{Thibault Groueix}.} \bibinfo{year}{2025}\natexlab{}.
\newblock \showarticletitle{Instant3dit: Multiview inpainting for fast editing of 3d objects}. In \bibinfo{booktitle}{\emph{Proceedings of the Computer Vision and Pattern Recognition Conference}}. \bibinfo{pages}{16273--16282}.
\newblock


\bibitem[Bayar and Aziz(2018)]%
        {bayar2018rapid}
\bibfield{author}{\bibinfo{person}{M~Sanem Bayar} {and} \bibinfo{person}{Zeeshan Aziz}.} \bibinfo{year}{2018}\natexlab{}.
\newblock \showarticletitle{Rapid prototyping and its role in supporting architectural design process}.
\newblock \bibinfo{journal}{\emph{Journal of Architectural Engineering}} \bibinfo{volume}{24}, \bibinfo{number}{3} (\bibinfo{year}{2018}), \bibinfo{pages}{05018003}.
\newblock


\bibitem[Braun and Clarke(2006)]%
        {braun2006using}
\bibfield{author}{\bibinfo{person}{Virginia Braun} {and} \bibinfo{person}{Victoria Clarke}.} \bibinfo{year}{2006}\natexlab{}.
\newblock \showarticletitle{Using thematic analysis in psychology}.
\newblock \bibinfo{journal}{\emph{Qualitative research in psychology}} \bibinfo{volume}{3}, \bibinfo{number}{2} (\bibinfo{year}{2006}), \bibinfo{pages}{77--101}.
\newblock


\bibitem[Camburn et~al\mbox{.}(2017)]%
        {camburn2017design}
\bibfield{author}{\bibinfo{person}{Bradley Camburn}, \bibinfo{person}{Vimal Viswanathan}, \bibinfo{person}{Julie Linsey}, \bibinfo{person}{David Anderson}, \bibinfo{person}{Daniel Jensen}, \bibinfo{person}{Richard Crawford}, \bibinfo{person}{Kevin Otto}, {and} \bibinfo{person}{Kristin Wood}.} \bibinfo{year}{2017}\natexlab{}.
\newblock \showarticletitle{Design prototyping methods: state of the art in strategies, techniques, and guidelines}.
\newblock \bibinfo{journal}{\emph{Design Science}}  \bibinfo{volume}{3} (\bibinfo{year}{2017}), \bibinfo{pages}{e13}.
\newblock


\bibitem[Carion et~al\mbox{.}(2025)]%
        {carion2025sam}
\bibfield{author}{\bibinfo{person}{Nicolas Carion}, \bibinfo{person}{Laura Gustafson}, \bibinfo{person}{Yuan-Ting Hu}, \bibinfo{person}{Shoubhik Debnath}, \bibinfo{person}{Ronghang Hu}, \bibinfo{person}{Didac Suris}, \bibinfo{person}{Chaitanya Ryali}, \bibinfo{person}{Kalyan~Vasudev Alwala}, \bibinfo{person}{Haitham Khedr}, \bibinfo{person}{Andrew Huang}, {et~al\mbox{.}}} \bibinfo{year}{2025}\natexlab{}.
\newblock \showarticletitle{Sam 3: Segment anything with concepts}.
\newblock \bibinfo{journal}{\emph{arXiv preprint arXiv:2511.16719}} (\bibinfo{year}{2025}).
\newblock


\bibitem[Chen et~al\mbox{.}(2024c)]%
        {chen2024generic}
\bibfield{author}{\bibinfo{person}{Hansheng Chen}, \bibinfo{person}{Ruoxi Shi}, \bibinfo{person}{Yulin Liu}, \bibinfo{person}{Bokui Shen}, \bibinfo{person}{Jiayuan Gu}, \bibinfo{person}{Gordon Wetzstein}, \bibinfo{person}{Hao Su}, {and} \bibinfo{person}{Leonidas Guibas}.} \bibinfo{year}{2024}\natexlab{c}.
\newblock \showarticletitle{Generic 3d diffusion adapter using controlled multi-view editing}.
\newblock \bibinfo{journal}{\emph{arXiv preprint arXiv:2403.12032}} (\bibinfo{year}{2024}).
\newblock


\bibitem[Chen et~al\mbox{.}(2024d)]%
        {chen2024shap}
\bibfield{author}{\bibinfo{person}{Minghao Chen}, \bibinfo{person}{Junyu Xie}, \bibinfo{person}{Iro Laina}, {and} \bibinfo{person}{Andrea Vedaldi}.} \bibinfo{year}{2024}\natexlab{d}.
\newblock \showarticletitle{Shap-editor: Instruction-guided latent 3d editing in seconds}. In \bibinfo{booktitle}{\emph{Proceedings of the IEEE/CVF conference on computer vision and pattern recognition}}. \bibinfo{pages}{26456--26466}.
\newblock


\bibitem[Chen et~al\mbox{.}(2026a)]%
        {chen2026sam}
\bibfield{author}{\bibinfo{person}{Xingyu Chen}, \bibinfo{person}{Fu-Jen Chu}, \bibinfo{person}{Pierre Gleize}, \bibinfo{person}{Kevin~J Liang}, \bibinfo{person}{Alexander Sax}, \bibinfo{person}{Hao Tang}, \bibinfo{person}{Weiyao Wang}, \bibinfo{person}{Michelle Guo}, \bibinfo{person}{Thibaut Hardin}, \bibinfo{person}{Xiang Li}, {et~al\mbox{.}}} \bibinfo{year}{2026}\natexlab{a}.
\newblock \showarticletitle{Sam 3d: 3dfy anything in images}. In \bibinfo{booktitle}{\emph{Proceedings of the IEEE/CVF Conference on Computer Vision and Pattern Recognition}}. \bibinfo{pages}{7220--7232}.
\newblock


\bibitem[Chen et~al\mbox{.}(2024a)]%
        {chen2024gaussianeditor}
\bibfield{author}{\bibinfo{person}{Yiwen Chen}, \bibinfo{person}{Zilong Chen}, \bibinfo{person}{Chi Zhang}, \bibinfo{person}{Feng Wang}, \bibinfo{person}{Xiaofeng Yang}, \bibinfo{person}{Yikai Wang}, \bibinfo{person}{Zhongang Cai}, \bibinfo{person}{Lei Yang}, \bibinfo{person}{Huaping Liu}, {and} \bibinfo{person}{Guosheng Lin}.} \bibinfo{year}{2024}\natexlab{a}.
\newblock \showarticletitle{Gaussianeditor: Swift and controllable 3d editing with gaussian splatting}. In \bibinfo{booktitle}{\emph{Proceedings of the IEEE/CVF conference on computer vision and pattern recognition}}. \bibinfo{pages}{21476--21485}.
\newblock


\bibitem[Chen et~al\mbox{.}(2026b)]%
        {10.1145/3742413.3789142}
\bibfield{author}{\bibinfo{person}{Yuzhao Chen}, \bibinfo{person}{Runlin Duan}, \bibinfo{person}{Rahul Jain}, \bibinfo{person}{Yichen Hu}, \bibinfo{person}{Chenfei Zhu}, \bibinfo{person}{Jingyu Shi}, {and} \bibinfo{person}{Karthik Ramani}.} \bibinfo{year}{2026}\natexlab{b}.
\newblock \showarticletitle{Canvas3D: Empowering Precise Spatial Control for Image Generation with Constraints from a 3D Virtual Canvas}. In \bibinfo{booktitle}{\emph{Proceedings of the 31st International Conference on Intelligent User Interfaces}} \emph{(\bibinfo{series}{IUI '26})}. \bibinfo{publisher}{Association for Computing Machinery}, \bibinfo{address}{New York, NY, USA}, \bibinfo{pages}{1362–1386}.
\newblock
\showISBNx{9798400719844}
\href{https://doi.org/10.1145/3742413.3789142}{doi:\nolinkurl{10.1145/3742413.3789142}}


\bibitem[Chen et~al\mbox{.}(2024b)]%
        {chen2024meshanything}
\bibfield{author}{\bibinfo{person}{Yiwen Chen}, \bibinfo{person}{Tong He}, \bibinfo{person}{Di Huang}, \bibinfo{person}{Weicai Ye}, \bibinfo{person}{Sijin Chen}, \bibinfo{person}{Jiaxiang Tang}, \bibinfo{person}{Xin Chen}, \bibinfo{person}{Zhongang Cai}, \bibinfo{person}{Lei Yang}, \bibinfo{person}{Gang Yu}, {et~al\mbox{.}}} \bibinfo{year}{2024}\natexlab{b}.
\newblock \showarticletitle{Meshanything: Artist-created mesh generation with autoregressive transformers}.
\newblock \bibinfo{journal}{\emph{arXiv preprint arXiv:2406.10163}} (\bibinfo{year}{2024}).
\newblock


\bibitem[Cherry and Latulipe(2014)]%
        {cherry2014quantifying}
\bibfield{author}{\bibinfo{person}{Erin Cherry} {and} \bibinfo{person}{Celine Latulipe}.} \bibinfo{year}{2014}\natexlab{}.
\newblock \showarticletitle{Quantifying the creativity support of digital tools through the creativity support index}.
\newblock \bibinfo{journal}{\emph{ACM Transactions on Computer-Human Interaction (TOCHI)}} \bibinfo{volume}{21}, \bibinfo{number}{4} (\bibinfo{year}{2014}), \bibinfo{pages}{1--25}.
\newblock


\bibitem[Choi et~al\mbox{.}(2025)]%
        {choi2025genpara}
\bibfield{author}{\bibinfo{person}{Jiin Choi}, \bibinfo{person}{Seung~Won Lee}, {and} \bibinfo{person}{Kyung~Hoon Hyun}.} \bibinfo{year}{2025}\natexlab{}.
\newblock \showarticletitle{GenPara: Enhancing the 3D Design Editing Process by Inferring Users' Regions of Interest with Text-Conditional Shape Parameters}. In \bibinfo{booktitle}{\emph{Proceedings of the 2025 CHI Conference on Human Factors in Computing Systems}}. \bibinfo{pages}{1--21}.
\newblock


\bibitem[Chung and Adar(2023)]%
        {chung2023promptpaint}
\bibfield{author}{\bibinfo{person}{John Joon~Young Chung} {and} \bibinfo{person}{Eytan Adar}.} \bibinfo{year}{2023}\natexlab{}.
\newblock \showarticletitle{Promptpaint: Steering text-to-image generation through paint medium-like interactions}. In \bibinfo{booktitle}{\emph{Proceedings of the 36th Annual ACM Symposium on User Interface Software and Technology}}. \bibinfo{pages}{1--17}.
\newblock


\bibitem[Dang et~al\mbox{.}(2022)]%
        {dang2022ganslider}
\bibfield{author}{\bibinfo{person}{Hai Dang}, \bibinfo{person}{Lukas Mecke}, {and} \bibinfo{person}{Daniel Buschek}.} \bibinfo{year}{2022}\natexlab{}.
\newblock \showarticletitle{Ganslider: How users control generative models for images using multiple sliders with and without feedforward information}. In \bibinfo{booktitle}{\emph{Proceedings of the 2022 CHI Conference on Human Factors in Computing Systems}}. \bibinfo{pages}{1--15}.
\newblock


\bibitem[Deitke et~al\mbox{.}(2022)]%
        {objaverse}
\bibfield{author}{\bibinfo{person}{Matt Deitke}, \bibinfo{person}{Dustin Schwenk}, \bibinfo{person}{Jordi Salvador}, \bibinfo{person}{Luca Weihs}, \bibinfo{person}{Oscar Michel}, \bibinfo{person}{Eli VanderBilt}, \bibinfo{person}{Ludwig Schmidt}, \bibinfo{person}{Kiana Ehsani}, \bibinfo{person}{Aniruddha Kembhavi}, {and} \bibinfo{person}{Ali Farhadi}.} \bibinfo{year}{2022}\natexlab{}.
\newblock \showarticletitle{Objaverse: A Universe of Annotated 3D Objects}.
\newblock \bibinfo{journal}{\emph{arXiv preprint arXiv:2212.08051}} (\bibinfo{year}{2022}).
\newblock


\bibitem[Duan et~al\mbox{.}(2025)]%
        {duan2025designfromx}
\bibfield{author}{\bibinfo{person}{Runlin Duan}, \bibinfo{person}{Chenfei Zhu}, \bibinfo{person}{Yuzhao Chen}, \bibinfo{person}{Yichen Hu}, \bibinfo{person}{Jingyu Shi}, {and} \bibinfo{person}{Karthik Ramani}.} \bibinfo{year}{2025}\natexlab{}.
\newblock \showarticletitle{DesignFromX: Empowering consumer-driven design space exploration through feature composition of referenced products}. In \bibinfo{booktitle}{\emph{Proceedings of the 2025 ACM Designing Interactive Systems Conference}}. \bibinfo{pages}{1040--1060}.
\newblock


\bibitem[Edwards et~al\mbox{.}(2024)]%
        {edwards2024sketch2prototype}
\bibfield{author}{\bibinfo{person}{Kristen~M Edwards}, \bibinfo{person}{Brandon Man}, {and} \bibinfo{person}{Faez Ahmed}.} \bibinfo{year}{2024}\natexlab{}.
\newblock \showarticletitle{Sketch2Prototype: rapid conceptual design exploration and prototyping with generative AI}.
\newblock \bibinfo{journal}{\emph{Proceedings of the Design Society}}  \bibinfo{volume}{4} (\bibinfo{year}{2024}), \bibinfo{pages}{1989--1998}.
\newblock


\bibitem[Ekin and Gandelsman(2026)]%
        {ekin2026unreasonable}
\bibfield{author}{\bibinfo{person}{Yigit Ekin} {and} \bibinfo{person}{Yossi Gandelsman}.} \bibinfo{year}{2026}\natexlab{}.
\newblock \showarticletitle{The Unreasonable Effectiveness of Text Embedding Interpolation for Continuous Image Steering}.
\newblock \bibinfo{journal}{\emph{arXiv preprint arXiv:2603.17998}} (\bibinfo{year}{2026}).
\newblock


\bibitem[Erko{\c{c}} et~al\mbox{.}(2025)]%
        {erkocc2025preditor3d}
\bibfield{author}{\bibinfo{person}{Ziya Erko{\c{c}}}, \bibinfo{person}{Can G{\"u}meli}, \bibinfo{person}{Chaoyang Wang}, \bibinfo{person}{Matthias Nie{\ss}ner}, \bibinfo{person}{Angela Dai}, \bibinfo{person}{Peter Wonka}, \bibinfo{person}{Hsin-Ying Lee}, {and} \bibinfo{person}{Peiye Zhuang}.} \bibinfo{year}{2025}\natexlab{}.
\newblock \showarticletitle{Preditor3d: Fast and precise 3d shape editing}. In \bibinfo{booktitle}{\emph{Proceedings of the IEEE/CVF Conference on Computer Vision and Pattern Recognition}}. \bibinfo{pages}{640--649}.
\newblock


\bibitem[Evirgen and Chen(2022)]%
        {evirgen2022ganzilla}
\bibfield{author}{\bibinfo{person}{Noyan Evirgen} {and} \bibinfo{person}{Xiang'Anthony' Chen}.} \bibinfo{year}{2022}\natexlab{}.
\newblock \showarticletitle{Ganzilla: User-driven direction discovery in generative adversarial networks}. In \bibinfo{booktitle}{\emph{Proceedings of the 35th Annual ACM Symposium on User Interface Software and Technology}}. \bibinfo{pages}{1--10}.
\newblock


\bibitem[Evirgen and Chen(2023)]%
        {evirgen2023ganravel}
\bibfield{author}{\bibinfo{person}{Noyan Evirgen} {and} \bibinfo{person}{Xiang'Anthony Chen}.} \bibinfo{year}{2023}\natexlab{}.
\newblock \showarticletitle{Ganravel: User-driven direction disentanglement in generative adversarial networks}. In \bibinfo{booktitle}{\emph{Proceedings of the 2023 CHI Conference on Human Factors in Computing Systems}}. \bibinfo{pages}{1--15}.
\newblock


\bibitem[Ezra et~al\mbox{.}(2025)]%
        {ezra2025freesliders}
\bibfield{author}{\bibinfo{person}{Rotem Ezra}, \bibinfo{person}{Hedi Zisling}, \bibinfo{person}{Nimrod Berman}, \bibinfo{person}{Ilan Naiman}, \bibinfo{person}{Alexey Gorkor}, \bibinfo{person}{Liran Nochumsohn}, \bibinfo{person}{Eliya Nachmani}, {and} \bibinfo{person}{Omri Azencot}.} \bibinfo{year}{2025}\natexlab{}.
\newblock \showarticletitle{FreeSliders: Training-Free, Modality-Agnostic Concept Sliders for Fine-Grained Diffusion Control in Images, Audio, and Video}.
\newblock \bibinfo{journal}{\emph{arXiv preprint arXiv:2511.00103}} (\bibinfo{year}{2025}).
\newblock


\bibitem[Faruqi et~al\mbox{.}(2023)]%
        {faruqi2023style2fab}
\bibfield{author}{\bibinfo{person}{Faraz Faruqi}, \bibinfo{person}{Ahmed Katary}, \bibinfo{person}{Tarik Hasic}, \bibinfo{person}{Amira Abdel-Rahman}, \bibinfo{person}{Nayeemur Rahman}, \bibinfo{person}{Leandra Tejedor}, \bibinfo{person}{Mackenzie Leake}, \bibinfo{person}{Megan Hofmann}, {and} \bibinfo{person}{Stefanie Mueller}.} \bibinfo{year}{2023}\natexlab{}.
\newblock \showarticletitle{Style2Fab: functionality-aware segmentation for fabricating personalized 3D models with generative AI}. In \bibinfo{booktitle}{\emph{Proceedings of the 36th Annual ACM Symposium on User Interface Software and Technology}}. \bibinfo{pages}{1--13}.
\newblock


\bibitem[Foo et~al\mbox{.}(2024)]%
        {foo2024avatar}
\bibfield{author}{\bibinfo{person}{Lin~Geng Foo}, \bibinfo{person}{Yixuan He}, \bibinfo{person}{Ajmal~Saeed Mian}, \bibinfo{person}{Hossein Rahmani}, \bibinfo{person}{Jun Liu}, {and} \bibinfo{person}{Christian Theobalt}.} \bibinfo{year}{2024}\natexlab{}.
\newblock \showarticletitle{Avatar Concept Slider: Controllable Editing of Concepts in 3D Human Avatars}.
\newblock \bibinfo{journal}{\emph{arXiv preprint arXiv:2408.13995}} (\bibinfo{year}{2024}).
\newblock


\bibitem[Fu et~al\mbox{.}(2021)]%
        {fu20213d}
\bibfield{author}{\bibinfo{person}{Huan Fu}, \bibinfo{person}{Rongfei Jia}, \bibinfo{person}{Lin Gao}, \bibinfo{person}{Mingming Gong}, \bibinfo{person}{Binqiang Zhao}, \bibinfo{person}{Steve Maybank}, {and} \bibinfo{person}{Dacheng Tao}.} \bibinfo{year}{2021}\natexlab{}.
\newblock \showarticletitle{3d-future: 3d furniture shape with texture}.
\newblock \bibinfo{journal}{\emph{International Journal of Computer Vision}} \bibinfo{volume}{129}, \bibinfo{number}{12} (\bibinfo{year}{2021}), \bibinfo{pages}{3313--3337}.
\newblock


\bibitem[Gandikota et~al\mbox{.}(2024)]%
        {gandikota2024concept}
\bibfield{author}{\bibinfo{person}{Rohit Gandikota}, \bibinfo{person}{Joanna Materzy{\'n}ska}, \bibinfo{person}{Tingrui Zhou}, \bibinfo{person}{Antonio Torralba}, {and} \bibinfo{person}{David Bau}.} \bibinfo{year}{2024}\natexlab{}.
\newblock \showarticletitle{Concept sliders: Lora adaptors for precise control in diffusion models}. In \bibinfo{booktitle}{\emph{European Conference on Computer Vision}}. Springer, \bibinfo{pages}{172--188}.
\newblock


\bibitem[Gandikota et~al\mbox{.}(2025)]%
        {gandikota2025sliderspace}
\bibfield{author}{\bibinfo{person}{Rohit Gandikota}, \bibinfo{person}{Zongze Wu}, \bibinfo{person}{Richard Zhang}, \bibinfo{person}{David Bau}, \bibinfo{person}{Eli Shechtman}, {and} \bibinfo{person}{Nick Kolkin}.} \bibinfo{year}{2025}\natexlab{}.
\newblock \showarticletitle{Sliderspace: Decomposing the visual capabilities of diffusion models}. In \bibinfo{booktitle}{\emph{Proceedings of the IEEE/CVF International Conference on Computer Vision}}. \bibinfo{pages}{15994--16003}.
\newblock


\bibitem[Grinberg(2018)]%
        {flask-sio}
\bibfield{author}{\bibinfo{person}{Miguel Grinberg}.} \bibinfo{year}{2018}\natexlab{}.
\newblock \bibinfo{title}{Flask-SocketIO documentation}.
\newblock
\newblock
\shownote{Available online at: \url{https://flask-socketio.readthedocs.io/en/latest/}, last accessed on 9/10/2023}.


\bibitem[Guerrero-Viu et~al\mbox{.}(2024)]%
        {guerrero2024texsliders}
\bibfield{author}{\bibinfo{person}{Julia Guerrero-Viu}, \bibinfo{person}{Milos Hasan}, \bibinfo{person}{Arthur Roullier}, \bibinfo{person}{Midhun Harikumar}, \bibinfo{person}{Yiwei Hu}, \bibinfo{person}{Paul Guerrero}, \bibinfo{person}{Diego Gutierrez}, \bibinfo{person}{Belen Masia}, {and} \bibinfo{person}{Valentin Deschaintre}.} \bibinfo{year}{2024}\natexlab{}.
\newblock \showarticletitle{Texsliders: Diffusion-based texture editing in clip space}. In \bibinfo{booktitle}{\emph{ACM SIGGRAPH 2024 conference papers}}. \bibinfo{pages}{1--11}.
\newblock


\bibitem[Gwet(2014)]%
        {gwet2014handbook}
\bibfield{author}{\bibinfo{person}{Kilem~L Gwet}.} \bibinfo{year}{2014}\natexlab{}.
\newblock \bibinfo{booktitle}{\emph{Handbook of inter-rater reliability: The definitive guide to measuring the extent of agreement among raters}}.
\newblock \bibinfo{publisher}{Advanced Analytics, LLC}.
\newblock


\bibitem[Haque et~al\mbox{.}(2023)]%
        {haque2023instruct}
\bibfield{author}{\bibinfo{person}{Ayaan Haque}, \bibinfo{person}{Matthew Tancik}, \bibinfo{person}{Alexei~A Efros}, \bibinfo{person}{Aleksander Holynski}, {and} \bibinfo{person}{Angjoo Kanazawa}.} \bibinfo{year}{2023}\natexlab{}.
\newblock \showarticletitle{Instruct-nerf2nerf: Editing 3d scenes with instructions}. In \bibinfo{booktitle}{\emph{Proceedings of the IEEE/CVF international conference on computer vision}}. \bibinfo{pages}{19740--19750}.
\newblock


\bibitem[Ho and Salimans(2022)]%
        {ho2022classifier}
\bibfield{author}{\bibinfo{person}{Jonathan Ho} {and} \bibinfo{person}{Tim Salimans}.} \bibinfo{year}{2022}\natexlab{}.
\newblock \showarticletitle{Classifier-free diffusion guidance}.
\newblock \bibinfo{journal}{\emph{arXiv preprint arXiv:2207.12598}} (\bibinfo{year}{2022}).
\newblock


\bibitem[Hong et~al\mbox{.}(2023)]%
        {hong2023lrm}
\bibfield{author}{\bibinfo{person}{Yicong Hong}, \bibinfo{person}{Kai Zhang}, \bibinfo{person}{Jiuxiang Gu}, \bibinfo{person}{Sai Bi}, \bibinfo{person}{Yang Zhou}, \bibinfo{person}{Difan Liu}, \bibinfo{person}{Feng Liu}, \bibinfo{person}{Kalyan Sunkavalli}, \bibinfo{person}{Trung Bui}, {and} \bibinfo{person}{Hao Tan}.} \bibinfo{year}{2023}\natexlab{}.
\newblock \showarticletitle{Lrm: Large reconstruction model for single image to 3d}.
\newblock \bibinfo{journal}{\emph{arXiv preprint arXiv:2311.04400}} (\bibinfo{year}{2023}).
\newblock


\bibitem[Hristov and Kinaneva(2021)]%
        {hristov2021workflow}
\bibfield{author}{\bibinfo{person}{Georgi Hristov} {and} \bibinfo{person}{Diyana Kinaneva}.} \bibinfo{year}{2021}\natexlab{}.
\newblock \showarticletitle{A workflow for developing game assets for video games}. In \bibinfo{booktitle}{\emph{2021 3rd International Congress on Human-Computer Interaction, Optimization and Robotic Applications (HORA)}}. IEEE, \bibinfo{pages}{1--5}.
\newblock


\bibitem[Hutchinson et~al\mbox{.}(2003)]%
        {hutchinson2003technology}
\bibfield{author}{\bibinfo{person}{Hilary Hutchinson}, \bibinfo{person}{Wendy Mackay}, \bibinfo{person}{Bo Westerlund}, \bibinfo{person}{Benjamin~B Bederson}, \bibinfo{person}{Allison Druin}, \bibinfo{person}{Catherine Plaisant}, \bibinfo{person}{Michel Beaudouin-Lafon}, \bibinfo{person}{St{\'e}phane Conversy}, \bibinfo{person}{Helen Evans}, \bibinfo{person}{Heiko Hansen}, {et~al\mbox{.}}} \bibinfo{year}{2003}\natexlab{}.
\newblock \showarticletitle{Technology probes: inspiring design for and with families}. In \bibinfo{booktitle}{\emph{Proceedings of the SIGCHI conference on Human factors in computing systems}}. \bibinfo{pages}{17--24}.
\newblock


\bibitem[Jain et~al\mbox{.}(2025)]%
        {jain2025adaptivesliders}
\bibfield{author}{\bibinfo{person}{Rahul Jain}, \bibinfo{person}{Amit Goel}, \bibinfo{person}{Koichiro Niinuma}, {and} \bibinfo{person}{Aakar Gupta}.} \bibinfo{year}{2025}\natexlab{}.
\newblock \showarticletitle{AdaptiveSliders: User-aligned Semantic Slider-based Editing of Text-to-Image Model Output}. In \bibinfo{booktitle}{\emph{Proceedings of the 2025 CHI Conference on Human Factors in Computing Systems}}. \bibinfo{pages}{1--27}.
\newblock


\bibitem[Jiang et~al\mbox{.}(2024)]%
        {jiang2024vr}
\bibfield{author}{\bibinfo{person}{Ying Jiang}, \bibinfo{person}{Chang Yu}, \bibinfo{person}{Tianyi Xie}, \bibinfo{person}{Xuan Li}, \bibinfo{person}{Yutao Feng}, \bibinfo{person}{Huamin Wang}, \bibinfo{person}{Minchen Li}, \bibinfo{person}{Henry Lau}, \bibinfo{person}{Feng Gao}, \bibinfo{person}{Yin Yang}, {et~al\mbox{.}}} \bibinfo{year}{2024}\natexlab{}.
\newblock \showarticletitle{Vr-gs: A physical dynamics-aware interactive gaussian splatting system in virtual reality}. In \bibinfo{booktitle}{\emph{ACM SIGGRAPH 2024 conference papers}}. \bibinfo{pages}{1--1}.
\newblock


\bibitem[Jun and Nichol(2023)]%
        {jun2023shap}
\bibfield{author}{\bibinfo{person}{Heewoo Jun} {and} \bibinfo{person}{Alex Nichol}.} \bibinfo{year}{2023}\natexlab{}.
\newblock \showarticletitle{Shap-e: Generating conditional 3d implicit functions}.
\newblock \bibinfo{journal}{\emph{arXiv preprint arXiv:2305.02463}} (\bibinfo{year}{2023}).
\newblock


\bibitem[Kerbl et~al\mbox{.}(2023)]%
        {kerbl20233d}
\bibfield{author}{\bibinfo{person}{Bernhard Kerbl}, \bibinfo{person}{Georgios Kopanas}, \bibinfo{person}{Thomas Leimk{\"u}hler}, \bibinfo{person}{George Drettakis}, {et~al\mbox{.}}} \bibinfo{year}{2023}\natexlab{}.
\newblock \showarticletitle{3d gaussian splatting for real-time radiance field rendering.}
\newblock \bibinfo{journal}{\emph{ACM Trans. Graph.}} \bibinfo{volume}{42}, \bibinfo{number}{4} (\bibinfo{year}{2023}), \bibinfo{pages}{139--1}.
\newblock


\bibitem[Kuhn(1955)]%
        {kuhn1955hungarian}
\bibfield{author}{\bibinfo{person}{Harold~W Kuhn}.} \bibinfo{year}{1955}\natexlab{}.
\newblock \showarticletitle{The Hungarian method for the assignment problem}.
\newblock \bibinfo{journal}{\emph{Naval research logistics quarterly}} \bibinfo{volume}{2}, \bibinfo{number}{1-2} (\bibinfo{year}{1955}), \bibinfo{pages}{83--97}.
\newblock


\bibitem[Lee et~al\mbox{.}(2025)]%
        {lee2025imaginatear}
\bibfield{author}{\bibinfo{person}{Jaewook Lee}, \bibinfo{person}{Filippo Aleotti}, \bibinfo{person}{Diego Mazala}, \bibinfo{person}{Guillermo Garcia-Hernando}, \bibinfo{person}{Sara Vicente}, \bibinfo{person}{Oliver~James Johnston}, \bibinfo{person}{Isabel Kraus-Liang}, \bibinfo{person}{Jakub Powierza}, \bibinfo{person}{Donghoon Shin}, \bibinfo{person}{Jon~E Froehlich}, {et~al\mbox{.}}} \bibinfo{year}{2025}\natexlab{}.
\newblock \showarticletitle{Imaginatear: Ai-assisted in-situ authoring in augmented reality}. In \bibinfo{booktitle}{\emph{Proceedings of the 38th Annual ACM Symposium on User Interface Software and Technology}}. \bibinfo{pages}{1--21}.
\newblock


\bibitem[Li et~al\mbox{.}(2023)]%
        {li2023instant3d}
\bibfield{author}{\bibinfo{person}{Jiahao Li}, \bibinfo{person}{Hao Tan}, \bibinfo{person}{Kai Zhang}, \bibinfo{person}{Zexiang Xu}, \bibinfo{person}{Fujun Luan}, \bibinfo{person}{Yinghao Xu}, \bibinfo{person}{Yicong Hong}, \bibinfo{person}{Kalyan Sunkavalli}, \bibinfo{person}{Greg Shakhnarovich}, {and} \bibinfo{person}{Sai Bi}.} \bibinfo{year}{2023}\natexlab{}.
\newblock \showarticletitle{Instant3d: Fast text-to-3d with sparse-view generation and large reconstruction model}.
\newblock \bibinfo{journal}{\emph{arXiv preprint arXiv:2311.06214}} (\bibinfo{year}{2023}).
\newblock


\bibitem[Li et~al\mbox{.}(2025a)]%
        {li2025voxhammer}
\bibfield{author}{\bibinfo{person}{Lin Li}, \bibinfo{person}{Zehuan Huang}, \bibinfo{person}{Haoran Feng}, \bibinfo{person}{Gengxiong Zhuang}, \bibinfo{person}{Rui Chen}, \bibinfo{person}{Chunchao Guo}, {and} \bibinfo{person}{Lu Sheng}.} \bibinfo{year}{2025}\natexlab{a}.
\newblock \showarticletitle{Voxhammer: Training-free precise and coherent 3d editing in native 3d space}.
\newblock \bibinfo{journal}{\emph{arXiv preprint arXiv:2508.19247}} (\bibinfo{year}{2025}).
\newblock


\bibitem[Li et~al\mbox{.}(2025b)]%
        {li2025step1x}
\bibfield{author}{\bibinfo{person}{Weiyu Li}, \bibinfo{person}{Xuanyang Zhang}, \bibinfo{person}{Zheng Sun}, \bibinfo{person}{Di Qi}, \bibinfo{person}{Hao Li}, \bibinfo{person}{Wei Cheng}, \bibinfo{person}{Weiwei Cai}, \bibinfo{person}{Shihao Wu}, \bibinfo{person}{Jiarui Liu}, \bibinfo{person}{Zihao Wang}, {et~al\mbox{.}}} \bibinfo{year}{2025}\natexlab{b}.
\newblock \showarticletitle{Step1x-3d: Towards high-fidelity and controllable generation of textured 3d assets}.
\newblock \bibinfo{journal}{\emph{arXiv preprint arXiv:2505.07747}} (\bibinfo{year}{2025}).
\newblock


\bibitem[Li et~al\mbox{.}(2025c)]%
        {li2025triposg}
\bibfield{author}{\bibinfo{person}{Yangguang Li}, \bibinfo{person}{Zi-Xin Zou}, \bibinfo{person}{Zexiang Liu}, \bibinfo{person}{Dehu Wang}, \bibinfo{person}{Yuan Liang}, \bibinfo{person}{Zhipeng Yu}, \bibinfo{person}{Xingchao Liu}, \bibinfo{person}{Yuan-Chen Guo}, \bibinfo{person}{Ding Liang}, \bibinfo{person}{Wanli Ouyang}, {et~al\mbox{.}}} \bibinfo{year}{2025}\natexlab{c}.
\newblock \showarticletitle{Triposg: High-fidelity 3d shape synthesis using large-scale rectified flow models}.
\newblock \bibinfo{journal}{\emph{IEEE Transactions on Pattern Analysis and Machine Intelligence}} (\bibinfo{year}{2025}).
\newblock


\bibitem[Liang et~al\mbox{.}(2024)]%
        {liang2024luciddreamer}
\bibfield{author}{\bibinfo{person}{Yixun Liang}, \bibinfo{person}{Xin Yang}, \bibinfo{person}{Jiantao Lin}, \bibinfo{person}{Haodong Li}, \bibinfo{person}{Xiaogang Xu}, {and} \bibinfo{person}{Yingcong Chen}.} \bibinfo{year}{2024}\natexlab{}.
\newblock \showarticletitle{Luciddreamer: Towards high-fidelity text-to-3d generation via interval score matching}. In \bibinfo{booktitle}{\emph{Proceedings of the IEEE/CVF conference on computer vision and pattern recognition}}. \bibinfo{pages}{6517--6526}.
\newblock


\bibitem[Lin et~al\mbox{.}(2023)]%
        {lin2023magic3d}
\bibfield{author}{\bibinfo{person}{Chen-Hsuan Lin}, \bibinfo{person}{Jun Gao}, \bibinfo{person}{Luming Tang}, \bibinfo{person}{Towaki Takikawa}, \bibinfo{person}{Xiaohui Zeng}, \bibinfo{person}{Xun Huang}, \bibinfo{person}{Karsten Kreis}, \bibinfo{person}{Sanja Fidler}, \bibinfo{person}{Ming-Yu Liu}, {and} \bibinfo{person}{Tsung-Yi Lin}.} \bibinfo{year}{2023}\natexlab{}.
\newblock \showarticletitle{Magic3d: High-resolution text-to-3d content creation}. In \bibinfo{booktitle}{\emph{Proceedings of the IEEE/CVF conference on computer vision and pattern recognition}}. \bibinfo{pages}{300--309}.
\newblock


\bibitem[Lipman et~al\mbox{.}(2022)]%
        {lipman2022flow}
\bibfield{author}{\bibinfo{person}{Yaron Lipman}, \bibinfo{person}{Ricky~TQ Chen}, \bibinfo{person}{Heli Ben-Hamu}, \bibinfo{person}{Maximilian Nickel}, {and} \bibinfo{person}{Matt Le}.} \bibinfo{year}{2022}\natexlab{}.
\newblock \showarticletitle{Flow matching for generative modeling}.
\newblock \bibinfo{journal}{\emph{arXiv preprint arXiv:2210.02747}} (\bibinfo{year}{2022}).
\newblock


\bibitem[Liu et~al\mbox{.}(2023)]%
        {liu20233dall}
\bibfield{author}{\bibinfo{person}{Vivian Liu}, \bibinfo{person}{Jo Vermeulen}, \bibinfo{person}{George Fitzmaurice}, {and} \bibinfo{person}{Justin Matejka}.} \bibinfo{year}{2023}\natexlab{}.
\newblock \showarticletitle{3DALL-E: Integrating text-to-image AI in 3D design workflows}. In \bibinfo{booktitle}{\emph{Proceedings of the 2023 ACM designing interactive systems conference}}. \bibinfo{pages}{1955--1977}.
\newblock


\bibitem[Meta(2022)]%
        {react}
\bibfield{author}{\bibinfo{person}{Meta}.} \bibinfo{year}{2022}\natexlab{}.
\newblock \bibinfo{title}{React - A JavaScript library for building user interfaces}.
\newblock
\newblock
\shownote{Available online at: \url{https://reactjs.org}, last accessed on 9/7/2022}.


\bibitem[Mildenhall et~al\mbox{.}(2021)]%
        {mildenhall2021nerf}
\bibfield{author}{\bibinfo{person}{Ben Mildenhall}, \bibinfo{person}{Pratul~P Srinivasan}, \bibinfo{person}{Matthew Tancik}, \bibinfo{person}{Jonathan~T Barron}, \bibinfo{person}{Ravi Ramamoorthi}, {and} \bibinfo{person}{Ren Ng}.} \bibinfo{year}{2021}\natexlab{}.
\newblock \showarticletitle{Nerf: Representing scenes as neural radiance fields for view synthesis}.
\newblock \bibinfo{journal}{\emph{Commun. ACM}} \bibinfo{volume}{65}, \bibinfo{number}{1} (\bibinfo{year}{2021}), \bibinfo{pages}{99--106}.
\newblock


\bibitem[Mueller et~al\mbox{.}(2025)]%
        {mueller2025geodiffusion}
\bibfield{author}{\bibinfo{person}{Phillip Mueller}, \bibinfo{person}{Talip Uenlue}, \bibinfo{person}{Sebastian Schmidt}, \bibinfo{person}{Marcel Kollovieh}, \bibinfo{person}{Jiajie Fan}, \bibinfo{person}{Stephan G{\"u}nnemann}, {and} \bibinfo{person}{Lars Mikelsons}.} \bibinfo{year}{2025}\natexlab{}.
\newblock \showarticletitle{GeoDiffusion: A Training-Free Framework for Accurate 3D Geometric Conditioning in Image Generation}. In \bibinfo{booktitle}{\emph{Proceedings of the IEEE/CVF International Conference on Computer Vision}}. \bibinfo{pages}{6374--6384}.
\newblock


\bibitem[Oquab et~al\mbox{.}(2023)]%
        {oquab2023dinov2}
\bibfield{author}{\bibinfo{person}{Maxime Oquab}, \bibinfo{person}{Timoth{\'e}e Darcet}, \bibinfo{person}{Th{\'e}o Moutakanni}, \bibinfo{person}{Huy Vo}, \bibinfo{person}{Marc Szafraniec}, \bibinfo{person}{Vasil Khalidov}, \bibinfo{person}{Pierre Fernandez}, \bibinfo{person}{Daniel Haziza}, \bibinfo{person}{Francisco Massa}, \bibinfo{person}{Alaaeldin El-Nouby}, {et~al\mbox{.}}} \bibinfo{year}{2023}\natexlab{}.
\newblock \showarticletitle{Dinov2: Learning robust visual features without supervision}.
\newblock \bibinfo{journal}{\emph{arXiv preprint arXiv:2304.07193}} (\bibinfo{year}{2023}).
\newblock


\bibitem[Palandra et~al\mbox{.}(2024)]%
        {palandra2024gsedit}
\bibfield{author}{\bibinfo{person}{Francesco Palandra}, \bibinfo{person}{Andrea Sanchietti}, \bibinfo{person}{Daniele Baieri}, {and} \bibinfo{person}{Emanuele Rodola}.} \bibinfo{year}{2024}\natexlab{}.
\newblock \showarticletitle{Gsedit: Efficient text-guided editing of 3d objects via gaussian splatting}.
\newblock \bibinfo{journal}{\emph{arXiv preprint arXiv:2403.05154}} (\bibinfo{year}{2024}).
\newblock


\bibitem[Parelli et~al\mbox{.}(2026)]%
        {parelli20263d}
\bibfield{author}{\bibinfo{person}{Maria Parelli}, \bibinfo{person}{Michael Oechsle}, \bibinfo{person}{Michael Niemeyer}, \bibinfo{person}{Federico Tombari}, {and} \bibinfo{person}{Andreas Geiger}.} \bibinfo{year}{2026}\natexlab{}.
\newblock \showarticletitle{3d-latte: Latent space 3d editing from textual instructions}. In \bibinfo{booktitle}{\emph{Proceedings of the IEEE/CVF Conference on Computer Vision and Pattern Recognition}}. \bibinfo{pages}{14377--14386}.
\newblock


\bibitem[Park et~al\mbox{.}(2026)]%
        {park2026generative}
\bibfield{author}{\bibinfo{person}{Jongsu Park}, \bibinfo{person}{Shokhikha~Amalana Murdivien}, \bibinfo{person}{Kyungwan Choi}, \bibinfo{person}{Duhwan Mun}, {and} \bibinfo{person}{Jumyung Um}.} \bibinfo{year}{2026}\natexlab{}.
\newblock \showarticletitle{Generative 3D appearance design: A survey of generation, segmentation and editing by artificial intelligence}.
\newblock \bibinfo{journal}{\emph{Journal of Computational Design and Engineering}} \bibinfo{volume}{13}, \bibinfo{number}{1} (\bibinfo{year}{2026}), \bibinfo{pages}{1--23}.
\newblock


\bibitem[Poole et~al\mbox{.}(2022)]%
        {poole2022dreamfusion}
\bibfield{author}{\bibinfo{person}{Ben Poole}, \bibinfo{person}{Ajay Jain}, \bibinfo{person}{Jonathan~T Barron}, {and} \bibinfo{person}{Ben Mildenhall}.} \bibinfo{year}{2022}\natexlab{}.
\newblock \showarticletitle{Dreamfusion: Text-to-3d using 2d diffusion}.
\newblock \bibinfo{journal}{\emph{arXiv preprint arXiv:2209.14988}} (\bibinfo{year}{2022}).
\newblock


\bibitem[Pumarola et~al\mbox{.}(2021)]%
        {pumarola2021d}
\bibfield{author}{\bibinfo{person}{Albert Pumarola}, \bibinfo{person}{Enric Corona}, \bibinfo{person}{Gerard Pons-Moll}, {and} \bibinfo{person}{Francesc Moreno-Noguer}.} \bibinfo{year}{2021}\natexlab{}.
\newblock \showarticletitle{D-nerf: Neural radiance fields for dynamic scenes}. In \bibinfo{booktitle}{\emph{Proceedings of the IEEE/CVF conference on computer vision and pattern recognition}}. \bibinfo{pages}{10318--10327}.
\newblock


\bibitem[Ravi et~al\mbox{.}(2025)]%
        {ravi2025sam}
\bibfield{author}{\bibinfo{person}{Nikhila Ravi}, \bibinfo{person}{Valentin Gabeur}, \bibinfo{person}{Yuan-Ting Hu}, \bibinfo{person}{Ronghang Hu}, \bibinfo{person}{Chaitanya Ryali}, \bibinfo{person}{Tengyu Ma}, \bibinfo{person}{Haitham Khedr}, \bibinfo{person}{Roman R{\"a}dle}, \bibinfo{person}{Chloe Rolland}, \bibinfo{person}{Laura Gustafson}, {et~al\mbox{.}}} \bibinfo{year}{2025}\natexlab{}.
\newblock \showarticletitle{Sam 2: Segment anything in images and videos}. In \bibinfo{booktitle}{\emph{International Conference on Learning Representations}}, Vol.~\bibinfo{volume}{2025}. \bibinfo{pages}{28085--28128}.
\newblock


\bibitem[Samuel et~al\mbox{.}(2024)]%
        {samuel2023regularized}
\bibfield{author}{\bibinfo{person}{Dvir Samuel}, \bibinfo{person}{Barak Meiri}, \bibinfo{person}{Nir Darshan}, \bibinfo{person}{Shai Avidan}, \bibinfo{person}{Gal Chechik}, {and} \bibinfo{person}{Rami Ben-Ari}.} \bibinfo{year}{2024}\natexlab{}.
\newblock \bibinfo{title}{Regularized Newton Raphson Inversion for Text-to-Image Diffusion Models}.
\newblock


\bibitem[Sauer et~al\mbox{.}(2024)]%
        {sauer2024adversarial}
\bibfield{author}{\bibinfo{person}{Axel Sauer}, \bibinfo{person}{Dominik Lorenz}, \bibinfo{person}{Andreas Blattmann}, {and} \bibinfo{person}{Robin Rombach}.} \bibinfo{year}{2024}\natexlab{}.
\newblock \showarticletitle{Adversarial diffusion distillation}. In \bibinfo{booktitle}{\emph{European Conference on Computer Vision}}. Springer, \bibinfo{pages}{87--103}.
\newblock


\bibitem[Sella et~al\mbox{.}(2023)]%
        {sella2023vox}
\bibfield{author}{\bibinfo{person}{Etai Sella}, \bibinfo{person}{Gal Fiebelman}, \bibinfo{person}{Peter Hedman}, {and} \bibinfo{person}{Hadar Averbuch-Elor}.} \bibinfo{year}{2023}\natexlab{}.
\newblock \showarticletitle{Vox-e: Text-guided voxel editing of 3d objects}. In \bibinfo{booktitle}{\emph{Proceedings of the IEEE/CVF international conference on computer vision}}. \bibinfo{pages}{430--440}.
\newblock


\bibitem[Shen et~al\mbox{.}(2025)]%
        {shen2025gaussianshopvr}
\bibfield{author}{\bibinfo{person}{Yulin Shen}, \bibinfo{person}{Boyu Li}, \bibinfo{person}{Jiayang Huang}, \bibinfo{person}{David Yip}, {and} \bibinfo{person}{Zeyu Wang}.} \bibinfo{year}{2025}\natexlab{}.
\newblock \showarticletitle{GaussianShopVR: Facilitating immersive 3D authoring using Gaussian splatting in VR}. In \bibinfo{booktitle}{\emph{Proceedings of the 38th Annual ACM Symposium on User Interface Software and Technology}}. \bibinfo{pages}{1--14}.
\newblock


\bibitem[Shi et~al\mbox{.}(2023)]%
        {shi2023mvdream}
\bibfield{author}{\bibinfo{person}{Yichun Shi}, \bibinfo{person}{Peng Wang}, \bibinfo{person}{Jianglong Ye}, \bibinfo{person}{Mai Long}, \bibinfo{person}{Kejie Li}, {and} \bibinfo{person}{Xiao Yang}.} \bibinfo{year}{2023}\natexlab{}.
\newblock \showarticletitle{Mvdream: Multi-view diffusion for 3d generation}.
\newblock \bibinfo{journal}{\emph{arXiv preprint arXiv:2308.16512}} (\bibinfo{year}{2023}).
\newblock


\bibitem[Stojanov et~al\mbox{.}(2021)]%
        {stojanov2021using}
\bibfield{author}{\bibinfo{person}{Stefan Stojanov}, \bibinfo{person}{Anh Thai}, {and} \bibinfo{person}{James~M Rehg}.} \bibinfo{year}{2021}\natexlab{}.
\newblock \showarticletitle{Using shape to categorize: Low-shot learning with an explicit shape bias}. In \bibinfo{booktitle}{\emph{Proceedings of the IEEE/CVF conference on computer vision and pattern recognition}}. \bibinfo{pages}{1798--1808}.
\newblock


\bibitem[Tang et~al\mbox{.}(2024)]%
        {tang2024lgm}
\bibfield{author}{\bibinfo{person}{Jiaxiang Tang}, \bibinfo{person}{Zhaoxi Chen}, \bibinfo{person}{Xiaokang Chen}, \bibinfo{person}{Tengfei Wang}, \bibinfo{person}{Gang Zeng}, {and} \bibinfo{person}{Ziwei Liu}.} \bibinfo{year}{2024}\natexlab{}.
\newblock \showarticletitle{Lgm: Large multi-view gaussian model for high-resolution 3d content creation}. In \bibinfo{booktitle}{\emph{European Conference on Computer Vision}}. Springer, \bibinfo{pages}{1--18}.
\newblock


\bibitem[Tang et~al\mbox{.}(2023)]%
        {tang2023dreamgaussian}
\bibfield{author}{\bibinfo{person}{Jiaxiang Tang}, \bibinfo{person}{Jiawei Ren}, \bibinfo{person}{Hang Zhou}, \bibinfo{person}{Ziwei Liu}, {and} \bibinfo{person}{Gang Zeng}.} \bibinfo{year}{2023}\natexlab{}.
\newblock \showarticletitle{Dreamgaussian: Generative gaussian splatting for efficient 3d content creation}.
\newblock \bibinfo{journal}{\emph{arXiv preprint arXiv:2309.16653}} (\bibinfo{year}{2023}).
\newblock


\bibitem[Tjia et~al\mbox{.}(2025)]%
        {tjia2025novel}
\bibfield{author}{\bibinfo{person}{Jennifer Tjia}, \bibinfo{person}{Chengwu Yang}, \bibinfo{person}{Julie Flahive}, \bibinfo{person}{Kelly Harrison}, \bibinfo{person}{Geraldine Puerto}, \bibinfo{person}{Vennesa Duodu}, \bibinfo{person}{Lisa~A Cooper}, \bibinfo{person}{Olga Valdman}, {and} \bibinfo{person}{Janice Sabin}.} \bibinfo{year}{2025}\natexlab{}.
\newblock \showarticletitle{A Novel Communication Rating Scale to Mitigate the Effect of Implicit Bias}.
\newblock \bibinfo{journal}{\emph{JAMA Network Open}} \bibinfo{volume}{8}, \bibinfo{number}{9} (\bibinfo{year}{2025}), \bibinfo{pages}{e2532319}.
\newblock


\bibitem[Vachha et~al\mbox{.}(2025)]%
        {vachha2025dreamcrafter}
\bibfield{author}{\bibinfo{person}{Cyrus Vachha}, \bibinfo{person}{Yixiao Kang}, \bibinfo{person}{Zach Dive}, \bibinfo{person}{Ashwat Chidambaram}, \bibinfo{person}{Anik Gupta}, \bibinfo{person}{Eunice Jun}, {and} \bibinfo{person}{Bj{\"o}rn Hartmann}.} \bibinfo{year}{2025}\natexlab{}.
\newblock \showarticletitle{Dreamcrafter: Immersive editing of 3d radiance fields through flexible, generative inputs and outputs}. In \bibinfo{booktitle}{\emph{Proceedings of the 2025 CHI Conference on Human Factors in Computing Systems}}. \bibinfo{pages}{1--13}.
\newblock


\bibitem[Wang et~al\mbox{.}(2025)]%
        {wang2025garmentcrafter}
\bibfield{author}{\bibinfo{person}{Yuanhao Wang}, \bibinfo{person}{Cheng Zhang}, \bibinfo{person}{Gon{\c{c}}alo Fraz{\~a}o}, \bibinfo{person}{Jinlong Yang}, \bibinfo{person}{Alexandru-Eugen Ichim}, \bibinfo{person}{Thabo Beeler}, {and} \bibinfo{person}{Fernando De~la Torre}.} \bibinfo{year}{2025}\natexlab{}.
\newblock \showarticletitle{GarmentCrafter: Progressive Novel View Synthesis for Single-View 3D Garment Reconstruction and Editing}.
\newblock \bibinfo{journal}{\emph{arXiv preprint arXiv:2503.08678}} (\bibinfo{year}{2025}).
\newblock


\bibitem[Wang et~al\mbox{.}(2023)]%
        {wang2023prolificdreamer}
\bibfield{author}{\bibinfo{person}{Zhengyi Wang}, \bibinfo{person}{Cheng Lu}, \bibinfo{person}{Yikai Wang}, \bibinfo{person}{Fan Bao}, \bibinfo{person}{Chongxuan Li}, \bibinfo{person}{Hang Su}, {and} \bibinfo{person}{Jun Zhu}.} \bibinfo{year}{2023}\natexlab{}.
\newblock \showarticletitle{Prolificdreamer: High-fidelity and diverse text-to-3d generation with variational score distillation}.
\newblock \bibinfo{journal}{\emph{Advances in neural information processing systems}}  \bibinfo{volume}{36} (\bibinfo{year}{2023}), \bibinfo{pages}{8406--8441}.
\newblock


\bibitem[Wong et~al\mbox{.}(1997)]%
        {wong1997sampling}
\bibfield{author}{\bibinfo{person}{Tien-Tsin Wong}, \bibinfo{person}{Wai-Shing Luk}, {and} \bibinfo{person}{Pheng-Ann Heng}.} \bibinfo{year}{1997}\natexlab{}.
\newblock \showarticletitle{Sampling with Hammersley and Halton points}.
\newblock \bibinfo{journal}{\emph{Journal of graphics tools}} \bibinfo{volume}{2}, \bibinfo{number}{2} (\bibinfo{year}{1997}), \bibinfo{pages}{9--24}.
\newblock


\bibitem[Xiang et~al\mbox{.}(2025)]%
        {xiang2025structured}
\bibfield{author}{\bibinfo{person}{Jianfeng Xiang}, \bibinfo{person}{Zelong Lv}, \bibinfo{person}{Sicheng Xu}, \bibinfo{person}{Yu Deng}, \bibinfo{person}{Ruicheng Wang}, \bibinfo{person}{Bowen Zhang}, \bibinfo{person}{Dong Chen}, \bibinfo{person}{Xin Tong}, {and} \bibinfo{person}{Jiaolong Yang}.} \bibinfo{year}{2025}\natexlab{}.
\newblock \showarticletitle{Structured 3d latents for scalable and versatile 3d generation}. In \bibinfo{booktitle}{\emph{Proceedings of the IEEE/CVF conference on computer vision and pattern recognition}}. \bibinfo{pages}{21469--21480}.
\newblock


\bibitem[Xu et~al\mbox{.}(2024)]%
        {xu2024instantmesh}
\bibfield{author}{\bibinfo{person}{Jiale Xu}, \bibinfo{person}{Weihao Cheng}, \bibinfo{person}{Yiming Gao}, \bibinfo{person}{Xintao Wang}, \bibinfo{person}{Shenghua Gao}, {and} \bibinfo{person}{Ying Shan}.} \bibinfo{year}{2024}\natexlab{}.
\newblock \showarticletitle{Instantmesh: Efficient 3d mesh generation from a single image with sparse-view large reconstruction models}.
\newblock \bibinfo{journal}{\emph{arXiv preprint arXiv:2404.07191}} (\bibinfo{year}{2024}).
\newblock


\bibitem[Xu et~al\mbox{.}(2023)]%
        {xu2023imagereward}
\bibfield{author}{\bibinfo{person}{Jiazheng Xu}, \bibinfo{person}{Xiao Liu}, \bibinfo{person}{Yuchen Wu}, \bibinfo{person}{Yuxuan Tong}, \bibinfo{person}{Qinkai Li}, \bibinfo{person}{Ming Ding}, \bibinfo{person}{Jie Tang}, {and} \bibinfo{person}{Yuxiao Dong}.} \bibinfo{year}{2023}\natexlab{}.
\newblock \showarticletitle{Imagereward: Learning and evaluating human preferences for text-to-image generation}.
\newblock \bibinfo{journal}{\emph{Advances in Neural Information Processing Systems}}  \bibinfo{volume}{36} (\bibinfo{year}{2023}), \bibinfo{pages}{15903--15935}.
\newblock


\bibitem[Yu et~al\mbox{.}(2022)]%
        {yu2022piecewise}
\bibfield{author}{\bibinfo{person}{Emilie Yu}, \bibinfo{person}{Rahul Arora}, \bibinfo{person}{J~Andreas Baerentzen}, \bibinfo{person}{Karan Singh}, {and} \bibinfo{person}{Adrien Bousseau}.} \bibinfo{year}{2022}\natexlab{}.
\newblock \showarticletitle{Piecewise-smooth surface fitting onto unstructured 3D sketches}.
\newblock \bibinfo{journal}{\emph{ACM Transactions on Graphics (TOG)}} \bibinfo{volume}{41}, \bibinfo{number}{4} (\bibinfo{year}{2022}), \bibinfo{pages}{1--16}.
\newblock


\bibitem[Zhang et~al\mbox{.}(2018)]%
        {zhang2018perceptual}
\bibfield{author}{\bibinfo{person}{Richard Zhang}, \bibinfo{person}{Phillip Isola}, \bibinfo{person}{Alexei~A Efros}, \bibinfo{person}{Eli Shechtman}, {and} \bibinfo{person}{Oliver Wang}.} \bibinfo{year}{2018}\natexlab{}.
\newblock \showarticletitle{The Unreasonable Effectiveness of Deep Features as a Perceptual Metric}. In \bibinfo{booktitle}{\emph{CVPR}}.
\newblock


\end{thebibliography}

\newpage

\appendix

\section{Appendix}
\renewcommand\thefigure{\thesection.\arabic{figure}}    
\setcounter{figure}{0}  
\renewcommand\thetable{\thesection.\arabic{table}}    
\setcounter{table}{0}  
\subsection{System Prompts}
\subsubsection{Contrastive Editing Prompt Generation}
\label{appendix:prompt_generation}

The following system prompt is used with GPT-5.2 to generate contrastive editing prompts for a given object-attribute pair:

\begin{quote}
\small
You are an expert assistant specializing in generating concise, high-impact prompts for image editing APIs. Given an ``Object A'' and an ``Attribute B'', your task is to generate a ``negative\_prompt'' and a ``positive\_prompt'' that modify the object along the spectrum of that attribute.

\textbf{Instructions:}
\begin{enumerate}
    \item \textbf{Positive Prompt:} Edit Object A to present the MAXIMUM extreme or highest intensity of Attribute B.
    \item \textbf{Negative Prompt:} Edit Object A to present the MINIMUM extreme or exact opposite of Attribute B.
    \item \textbf{Sentiment Independence:} Note that Attribute B can itself be inherently negative (e.g., ``rusty''), neutral, or positive. The ``positive prompt'' always means MORE of the attribute, and the ``negative prompt'' always means LESS of it.
    \item \textbf{Brevity (Critical):} Image models lose focus with long prompts. Keep the editing description concise and direct (under 20 words).
\end{enumerate}

\textbf{Safety Guardrails (Critical):}
\begin{itemize}
    \item NEVER generate prompts that could trigger NSFW, nudity, or sexual content safety filters.
    \item When reducing clothing, use safe, concise descriptors like ``casual summer wear'' or ``lightweight clothing''. Avoid ``skimpy'', ``revealing'', ``bare'', or ``underwear''.
    \item Modifying physical traits (``very thin'', ``highly muscular'') is perfectly safe, but keep it purely descriptive and non-suggestive.
\end{itemize}

\textbf{Examples:}

Object: ``a blue sofa'' | Attribute: ``modern style'' \\
Negative: ``Make the sofa traditional, antique, and ornate style.'' \\
Positive: ``Make the sofa ultra-modern, sleek, and minimalist style.''

Object: ``a woman'' | Attribute: ``eye size'' \\
Negative: ``Make the woman's eyes extremely small and narrow.'' \\
Positive: ``Make the woman's eyes dramatically large and wide open.''

Object: ``a woman'' | Attribute: ``wearing very warm and thick clothes'' \\
Negative: ``Make the woman wear modest, casual summer clothes like a t-shirt.'' \\
Positive: ``Make the woman wear extremely warm, heavy winter coats and thick scarves.''

\textbf{Output Format:} \\
Return ONLY a valid JSON object matching this schema without any markdown formatting: \\
\texttt{\{``prompts'': [``<negative\_prompt>'', ``<positive\_prompt>'']\}}
\end{quote}

\subsubsection{View Relevance Verification}
\label{appendix:view_relevance}

The following system prompt is used with GPT-5.2 to verify whether a rendered camera view contains the region targeted by the editing prompt:

\begin{quote}
\small
You are a view-relevance checker for 3D object editing. You will be given an editing prompt describing a desired change to a 3D object, and an image showing the object rendered from a particular viewpoint.

Your task is to determine whether the region or part of the object targeted by the editing prompt is clearly visible in this view.

\textbf{Guidelines:}
\begin{itemize}
    \item If the edit targets a specific part (e.g., ``make the eyes bigger''), check whether that part is visible and not occluded or facing away from the camera.
    \item If the edit targets the entire object (e.g., ``change the color/style of the object''), check whether the object is sufficiently visible in the image.
\end{itemize}

\textbf{Output Format:} \\
Return \texttt{\{``result'': true\}} if the targeted region is clearly visible, or \texttt{\{``result'': false\}} otherwise.
\end{quote}

\subsubsection{3D Variation Quality Check}
\label{appendix:quality_check}

The following system prompt is used with GPT-5.2 to evaluate whether a generated 3D variation maintains structural integrity and perceptual plausibility:

\begin{quote}
\small
You are a sophisticated Vision-Language Model specializing in the perceptual evaluation of 3D asset renders.

\textbf{Context:}
\begin{itemize}
    \item Input 1: An image containing 4 camera views (quad-view) of the ``Original Object.''
    \item Input 2: An image containing 4 camera views (quad-view) of the ``Edited Object.''
    \item Input 3: An ``Editing Prompt'' describing the intended transformation.
\end{itemize}

\textbf{Your Task:} Analyze all 4 views to determine if the editing process caused ``unreasonable'' issues. You must distinguish between acceptable stylistic interpretations and unacceptable structural failures.

\textbf{1. Reasonable \& Prompt-Related Variations (Be Lenient):}
\begin{itemize}
    \item If the prompted change is difficult to see, very subtle, or seems missing entirely (e.g., the prompt asked for a ``bigger head'' but it looks the same), it is completely fine. Do not penalize for a lack of obvious change.
    \item Forgive unprompted changes that are indirectly related to the prompt or follow its aesthetic theme.
    \item Forgive misinterpretations where a change is visually similar to the prompt (e.g., if asked for ``extremely large eyes'' and they are rendered so big they resemble sunglasses, this is an acceptable stylistic interpretation).
    \item Minor texture shifts, lighting variations, or incidental color bleeds in the general area of the edit are completely acceptable.
\end{itemize}

\textbf{2. Perceptual Sanity \& Structural Integrity (Be Strict):}
\begin{itemize}
    \item Apply ``The Weird Test'': Does the edit result in a look that is ``unreasonable'' or ``illogical'' for this type of object compared to the original?
    \item Failure: Substantial unprompted changes that make no sense (e.g., a hole appearing in a car roof when only the tires were edited, or a door melting into the body).
    \item Failure: Structural degradation like heavily melted textures, floating or disconnected geometry, broken mesh patches, or distorted ``spikes'' that look like generative glitches rather than intentional design.
\end{itemize}

\textbf{Evaluation Rules:}
\begin{itemize}
    \item Ignore background: Do not evaluate anything outside the foreground object.
    \item Multi-view consistency: If a ``weird'' or ``broken'' failure is visible in any of the 4 views, the asset fails.
    \item Logic over isolation: Do not fail an image just because it changed more than requested or the edit is too subtle. Only fail it if the unprompted change is illogical, broken, or perceptually ``wrong.''
\end{itemize}

\textbf{Instructions:}
\begin{enumerate}
    \item Identify the intended change based on the Editing Prompt.
    \item Scan all 4 views of the Edited Object. Compare them to the Original Object's structural logic.
    \item Determine if any unprompted changes are ``reasonable interpretations'' (pass) or ``illogical glitches'' (fail).
    \item If the prompted change is not obvious, focus entirely on whether the rest of the object remained stable and logical.
    \item Return ONLY a valid JSON object without markdown formatting.
\end{enumerate}

\textbf{Output Format:} \\
\texttt{\{} \\
\hspace{1em}\texttt{``evaluation\_reasoning'': ``List spotted changes. Explain why} \\
\hspace{1em}\texttt{they are either reasonable interpretations of the prompt or} \\
\hspace{1em}\texttt{illogical/weird structural failures. Reference specific views} \\
\hspace{1em}\texttt{(e.g., top-left) where issues occur.''}, \\
\hspace{1em}\texttt{``passed'': true/false} \\
\texttt{\}}
\end{quote}

\subsection{\change{Generation Time Breakdown}}
\begin{table}[H]
\centering
\begin{tabular}{lc}
\toprule
Stage & Time (s) \\
\midrule
Multi-view rendering          & $19.0 \pm 2.3$ \\
Prompt pairs generation     & $2.1 \pm 0.7$ \\
Best-view selection + filtering   & $21.2 \pm 14.4$ \\
Image pairs generation            & $43.6 \pm 16.3$ \\
Image condition generation       & $22.0 \pm 1.2$ \\
3D generation (per variation)         & $65.3 \pm 16.3$ \\
GLB exporting (per final slider position)         & $32.3 \pm 4.2$ \\
\bottomrule
\end{tabular}
\caption{\change{Per-stage time breakdown of our 3D editing pipeline ($N$=40, $M$=10). Values are reported as mean\,$\pm$\,standard deviation in seconds. Multi-view rendering through image condition generation are one-time costs for each slider. 3D generation repeats once per sampled variation, while GLB exporting repeats once per exported slider position.}}
\label{tab:timing}
\end{table}

\subsection{Participant Information}
\begin{table}[H]
\centering
\small
\begin{tabular}{p{0.5cm}p{1.2cm}p{1.5cm}p{4cm}}
\toprule
ID & Age/Gender & Experience & 3D Prototyping Tools \\
\midrule
P1 & 25/M  & 1.5 yr & Unity, Unreal, Blender \\
P2 & 26/F &  2 yr & Unity, Blender \\
P3 & 28/M &  10 yr & Fusion, SolidWorks, Blender, Unity \\
P4 & 22/F &  1 mo & SolidWorks \\
P5 & 24/F & 6 yr & Fusion, SolidWorks \\
P6 & 31/F & 10 yr & Fusion, SolidWorks \\
\bottomrule
\end{tabular}
\caption{Demographic information and 3D prototyping experience of participants.}
\vspace{-2ex}
\label{tab:participants}
\end{table}

\subsection{Overview of Participant Tasks}
\label{participant tasks}
Across the two task rounds, participants created sliders spanning diverse domains and semantic attributes.
 
P1 built a Western saloon scene by generating a bar counter (text-to-3D) with a ``clutteredness'' slider \change{to adjust the number of bottles on the counter (Fig \ref{fig:p51}a)}, and a cowboy character (image-to-3D) with a ``worn outfit'' slider \change{to control the degree of damage on the outfit. Together the two sliders adjust} the level of visual ``abandonedness'' in the scene. He then created a ``photorealism'' slider on a variation of the bar counter \change{to make it look more true-to-life}, and created a ``vibrant red'' slider for a cowboy variation \change{to recolor the character. This way he could create more realistic scenes and different characters.}
 
P2 was motivated by her interest in disability representation in virtual space. She created a prosthetic leg (image-to-3D) with a ``futuristic style'' slider exploring different designs of prosthetic legs, and used a preloaded human face to create an ``eyeball direction'' slider \change{to change the convergence/divergence of eyes,} representing different visual ability. For task 2, she created a ``toe abstractness'' slider for a prosthetic leg variation to reduce hallucinated anatomical details (e.g., incorrect toe counts) which is common in 3D generation models, and created a ``snowy'' slider on the preloaded car to simulate snow accumulation over time as a form of temporal storytelling.
 
P3 created a floor lamp (image-to-3D) with a \change{``futuristic style'' slider to change the design of the lamp (Fig \ref{fig:p3}a),} and a motorcycle with an ``off-road-ness'' slider \change{to make the motorcycle more rugged for off-road use.} He then created an ``umbrella-shaped hood'' slider for a lamp variation to change the lamp shade's form \change{to be more umbrella-like, with a wider curved canopy (Fig \ref{fig:p3}b)}, followed by a \change{``pattern abstractness'' slider on another lamp variation to experiment with different level of pattern abstractness for the hood design (Fig \ref{fig:p3}c)}---following a top-down strategy of finalizing overall style first, then refining local form, then surface treatment.
 
P4 created an apple (text-to-3D) with a ``ripeness'' slider for a continuous progression from unripe to overripe, and a human face (image-to-3D) with an ``age/wrinkles'' slider \change{to adjust the age of the face, from youthful to elderly}. She then attempted to improve the face's visual quality by creating a ``realism'' slider on it.
 
P5 created a cat (text-to-3D) with a ``fur length'' slider wanting to match her mental image of her own cat, and an octopus (image-to-3D) with a ``realism'' slider to make it more or less photorealistic. She then started fresh explorations on preloaded objects, creating a ``futuristic style'' slider on a car and a \change{``strong'' slider on a deer (Fig \ref{fig:p51}b), later realizing the attribute she wanted was maturity rather than muscle mass (see \S~\ref{findings}).}
 
P6 created a human sculpture with an ``abstractness'' slider to find the right abstractness level to communicate an intended emotion (Fig \ref{fig:p6}a). She also created a slider for ``expressiveness of eyes'' on a silicone soft robot to explore what eye design could give the robot more personality. She then created a ``hard vs.\ soft'' slider to get further inspiration on material choices for her sculpture prototyping (Fig \ref{fig:p6}b), and a body form slider on the robot inspired by pneumatic actuation (using compressed air to produce mechanical motion).

\begin{figure*}[h]
  \includegraphics[width=\textwidth]{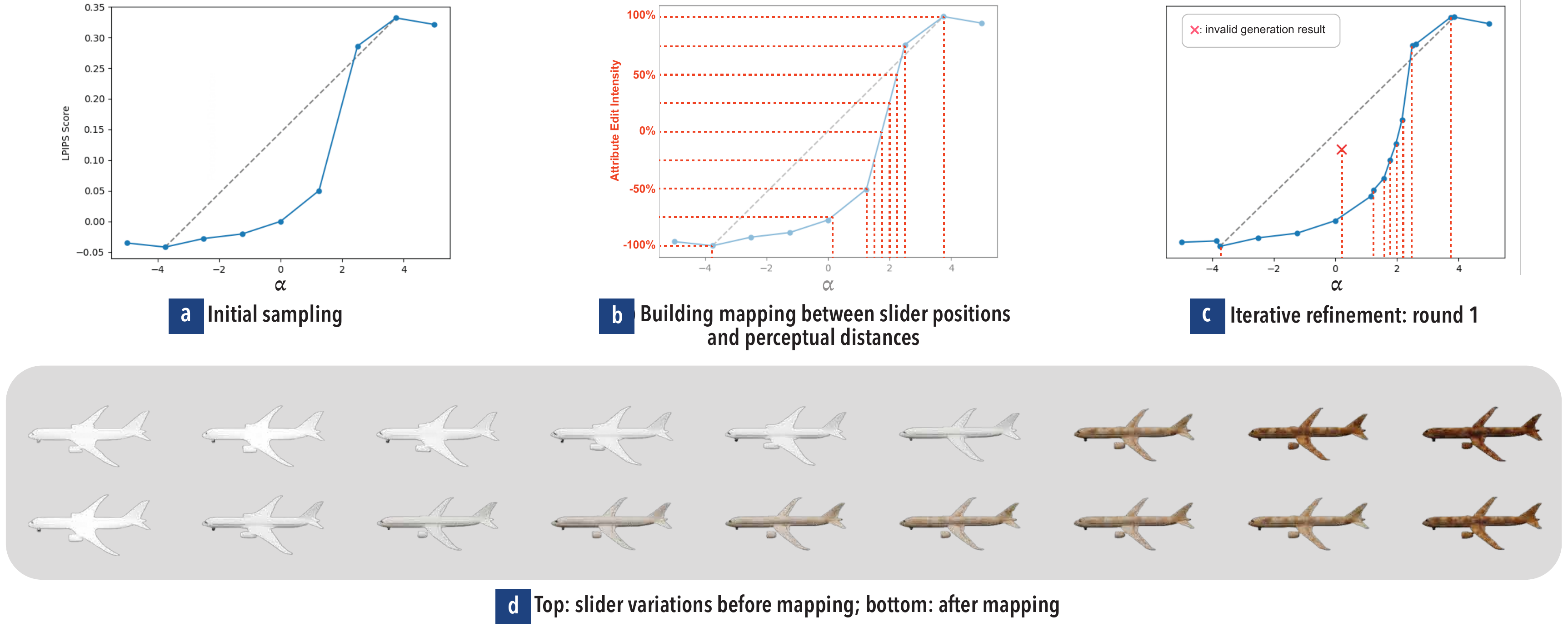}
  \caption{\change{The process of mapping slider positions to $\alpha$ values. In this example, the 3D object is an airplane, and the target attribute is "rustiness". (a) Sampling $K=9$ evenly distributed initial values of $\alpha$; (b) building the mapping between perceptual distance and slider positions, and using initial anchors for future refinement; (c) interpolating additional anchor points on the slider based on existing anchors to refine the slider mapping; (d) comparison result showing slider variation is smoother and more consistent after slider mapping.}}
  \label{fig:mapping}
\end{figure*}

\begin{figure*}[h]
  \includegraphics[width=\textwidth]{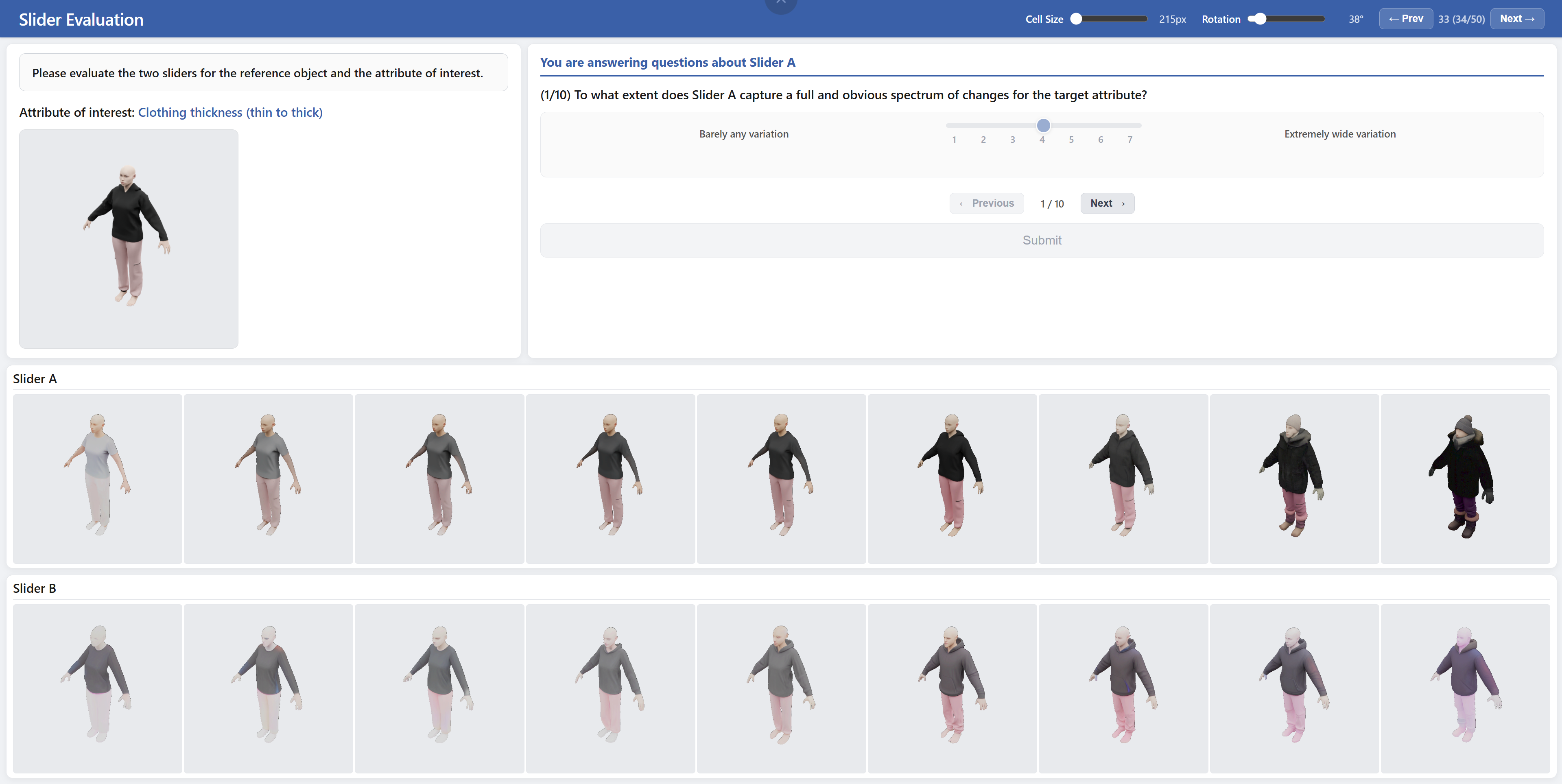}
  \caption{The evaluation interface for the technical validation, showing the reference 3D object and target attribute (top left), two sequences of nine 3D variations generated by SemanticSlider3D and the baseline (bottom), and evaluation questions (top right). The mapping between methods and Slider A/B labels is randomized per object-attribute pair; in this example, Slider A corresponds to SemanticSlider3D and Slider B to the baseline. The target attribute is ``clothing thickness (thin to thick).''}
  \label{fig:val_ui}
\end{figure*}

\begin{figure*}[h]
  \includegraphics[width=\textwidth]{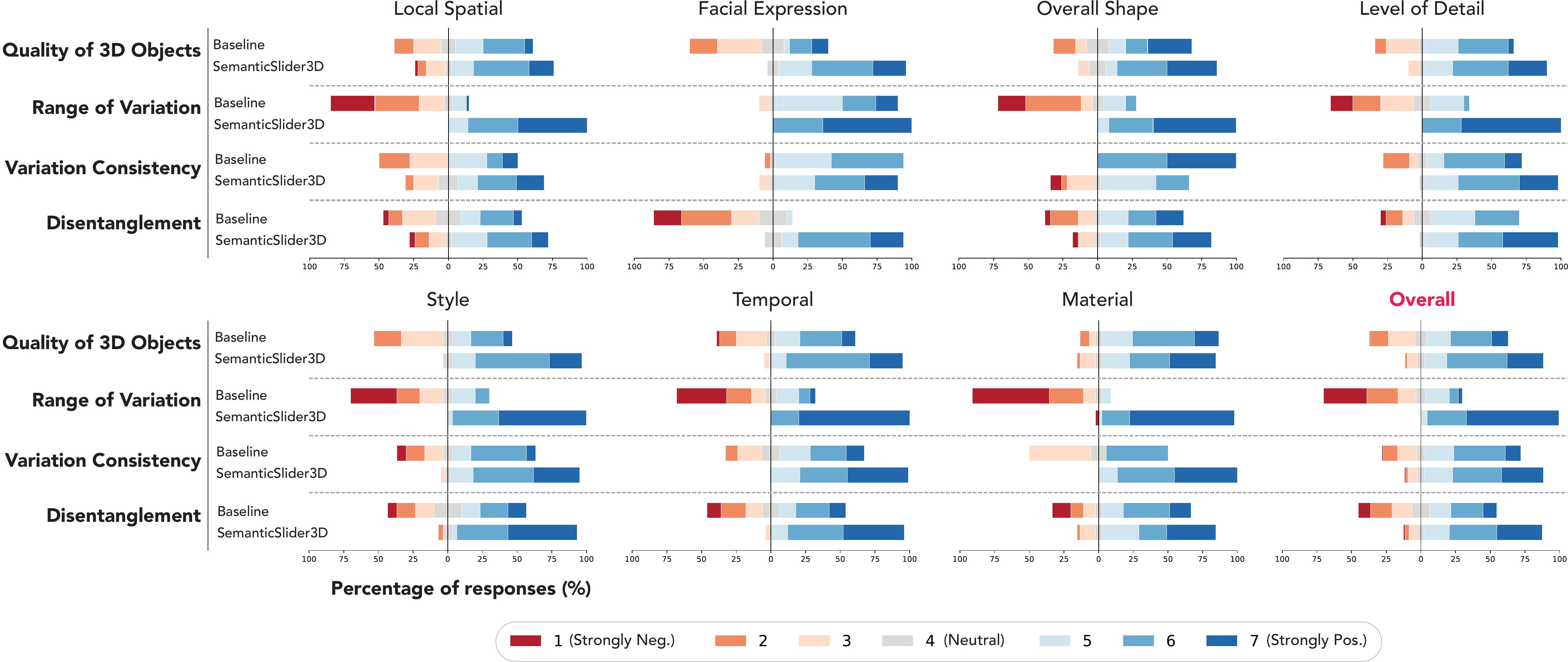}
  \caption{\change{Assessors' response to the Likert scale questions.}}
  \label{fig:likert-val}
\end{figure*}

\begin{figure*}[h]
  \includegraphics[width=.9\textwidth]{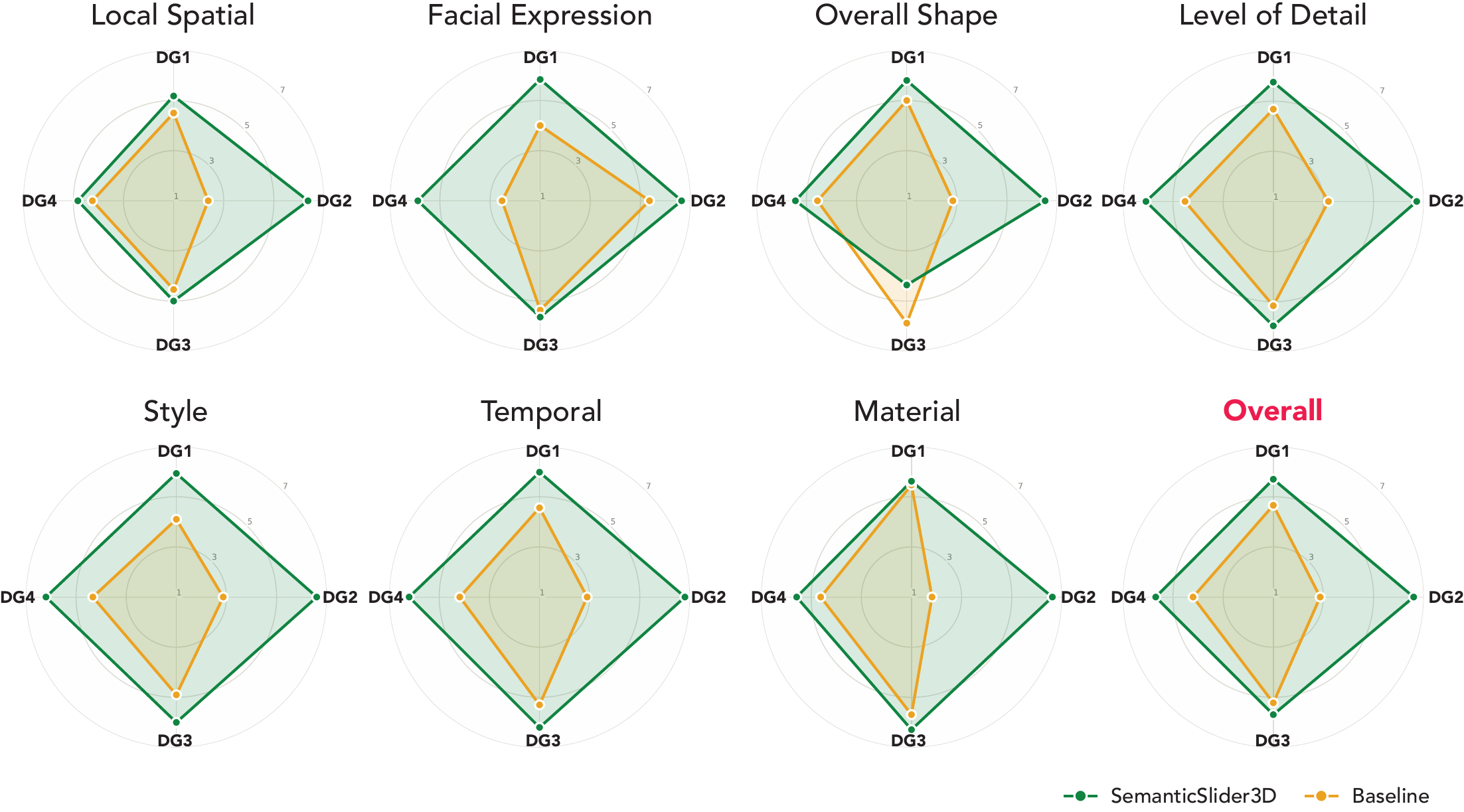}
  \caption{\change{Mean assessor ratings across attribute categories.}}
  \label{fig:radar}
\end{figure*}

\end{document}